\documentclass[fleqn,usenatbib]{mnras}

\usepackage{newtxtext,newtxmath}

\usepackage[T1]{fontenc}
\usepackage{ae,aecompl}

\usepackage{graphicx}	

\title[The {\em Spitzer} DeepDrill Survey]{A {\em Spitzer} survey of Deep Drilling Fields to be targeted by the Vera C.\ Rubin Observatory Legacy Survey of Space and Time}

\author[M. Lacy et al.]{
M.\ Lacy,$^{1}$
J.A.\ Surace,$^{2}$
D.\ Farrah,$^{3,4}$
K.\ Nyland,$^{5}$
J.\ Afonso,$^{6,7}$
W.N.\ Brandt,$^{8,9,10}$
\newauthor{D.L.\ Clements,$^{11}$ C.D.P.\ Lagos,$^{12,13,14}$ C.\ Maraston,$^{15}$ J.\ Pforr,$^{16}$ A.\ Sajina,$^{17}$} 
\newauthor{M.\ Sako,$^{18}$ M.\ Vaccari,$^{19,20}$ G.\ Wilson,$^{21}$ D.R.\ Ballantyne,$^{22}$ W.A.\ Barkhouse,$^{23}$}
\newauthor{R.\ Brunner,$^{18}$ R.\ Cane,$^{18}$ T.E.\ Clarke,$^{24}$ M.\ Cooper,$^{25}$, A.\ Cooray,$^{25}$}
\newauthor{G.\ Covone,$^{26}$ C.\ D'Andrea,$^{18}$ A.E.\ Evrard,$^{27,28}$ H.C.\ Ferguson,$^{29}$ J.\ Frieman,$^{30,31}$}
\newauthor{V. Gonzalez-Perez,$^{15,32}$ R.\ Gupta,$^{18}$ E.\ Hatziminaoglou,$^{33}$}
\newauthor{J.\ Huang,$^{34,35,36}$
P.\ Jagannathan,$^{37}$ M.J.\ Jarvis,$^{19,38}$ K.M.\ Jones,$^{39}$ A.\ Kimball,$^{37}$}
\newauthor{C.\ Lidman,$^{40}$, L.\ Lubin$^{41}$ L.\ Marchetti,$^{42,19,20}$P.\ Martini,$^{43,44}$R.G.\ McMahon,$^{45}$}
\newauthor{S.\ Mei,$^{46,47}$ H.\ Messias,$^{48}$ E.J.\ Murphy,$^{1}$ J.A.\ Newman,$^{49}$ R.\ Nichol,$^{15}$} 
\newauthor{R.P.\ Norris,$^{50}$ S.\ Oliver,$^{51}$ I.\ Perez-Fournon,$^{52,53}$ W.M.\ Peters,$^{24} $M.\ Pierre,$^{54}$}
\newauthor{E.\ Polisensky,$^{24}$ G.T.\ Richards,$^{55}$ S.E.\ Ridgway,$^{56}$ H.J.A.\ R\"{o}ttgering,$^{57}$ N.\ Seymour,$^{58}$}
\newauthor{R.\ Shirley,$^{51,52}$ R.\ Somerville,$^{59}$ 
M.A.\ Strauss,$^{60}$ N.\ Suntzeff,$^{61}$} \newauthor{P.A.\ Thorman,$^{62}$ E. van Kampen,$^{33}$ A.\ Verma,$^{38}$ R.\ Wechsler,$^{63}$ W.M.\ Wood-Vasey$^{64}$}}

\date{}

\pubyear{2019}

\begin{document}
\label{firstpage}
\pagerange{\pageref{firstpage}--\pageref{lastpage}}
\maketitle

\begin{abstract}

The Vera C.\ Rubin Observatory Legacy Survey of Space and Time (LSST) will observe several Deep Drilling Fields (DDFs) to a greater depth and with a more rapid cadence than the main survey. In this paper, we describe the ``DeepDrill'' survey, which used the  {\em Spitzer Space Telescope} Infrared Array Camera (IRAC) to observe three of the four currently defined DDFs in two bands, centered on 3.6 $\mu$m and 4.5 $\mu$m. These observations expand the area which was covered by an earlier set of observations in these three fields by the {\em Spitzer} Extragalactic Representative Volume Survey (SERVS). The combined DeepDrill and SERVS data cover the footprints of the LSST DDFs in the Extended Chandra Deep Field--South field (ECDFS), the ELAIS-S1 field (ES1), and the {\em XMM}-Large-Scale Structure Survey field ({\em XMM}-LSS). The observations reach an approximate $5\sigma$ point-source depth of 2~$\mu$Jy (corresponding to an AB magnitude of 23.1; sufficient to detect a 10$^{11} M_{\odot}$ galaxy out to $z\approx 5$) in each of the two bands over a total area of $\approx 29\,$deg$^2$. The dual-band catalogues contain a total of 2.35 million sources. In this paper we describe the observations and data products from the survey, and an overview of the properties of galaxies in the survey. We compare the source counts to predictions from the {\sc SHARK} semi-analytic model of galaxy formation. We also identify a population of sources with extremely red ([3.6]$-$[4.5] $>1.2$) colours which we show mostly consists of highly-obscured active galactic nuclei.  
\end{abstract}

\begin{keywords}
surveys -- infrared: general -- infrared:galaxies -- catalogues
\end{keywords}



\section{Introduction}
Surveys by the {\em Spitzer Space Telescope} have proved extremely valuable for finding and characterizing distant galaxies. The redshifting of the peak of stellar emission at $1.6\,\mu$m into the {\em Spitzer} bands makes them especially sensitive to high-redshift galaxies \citep[e.g.][]{2007A&A...476..151B,2015ApJ...803...11S,2019ApJ...880L..14C}. {\em Spitzer} data thus provide a very useful complement to deep surveys in the optical, where the surface density of galaxies is higher, but intrinsically luminous, high-redshift galaxies that are either quiescent or dust-reddened can be outnumbered by lower-redshift, lower luminosity blue galaxies. 
The Vera C.\ Rubin Observatory Legacy Survey of Space and Time \citep[LSST;][]{2019ApJ...873..111I} will observe the Southern sky in six optical bands ($u,\, g,\, r,\,i,\, z$ and  $y$) in about 800 passes (summed over all bands) over ten years, to a co-added 5$\sigma$ depth of $AB\approx 24.4-27.1$, depending on band. Within the survey area, there will be several Deep Drilling Fields (DDFs) where observations are repeated more frequently, resulting in both a better sampled cadence, and a deeper coadded final image \citep[$AB\approx 26.2-28.7$, depending on band;][]{2018arXiv181106542B,2018arXiv181200516S}. The DDFs will thus become important reference fields for both time domain and ultra-deep imaging studies. 

We therefore proposed to observe the DDFs that had already been defined by the LSST team
in the near-infrared with the {\em Spitzer Space Telescope} during its post-cryogenic mission (after the liquid helium cryogen supply for the telescope was exhausted in May 2009). Although only the two shortest wavelength bands of the Infrared Array Camera \citep[IRAC;][] {2004ApJS..154...10F,2010SPIE.7731E..0NC}, at 3.6 and 4.5~$\mu$m, continued in operation following the exhaustion of the cryogenic coolant in {\em Spitzer} in 2009, their sensitivity was almost unchanged, as was the optical behaviour of the telescope and instrument. 

\begin{figure}
    \includegraphics[scale=0.6]{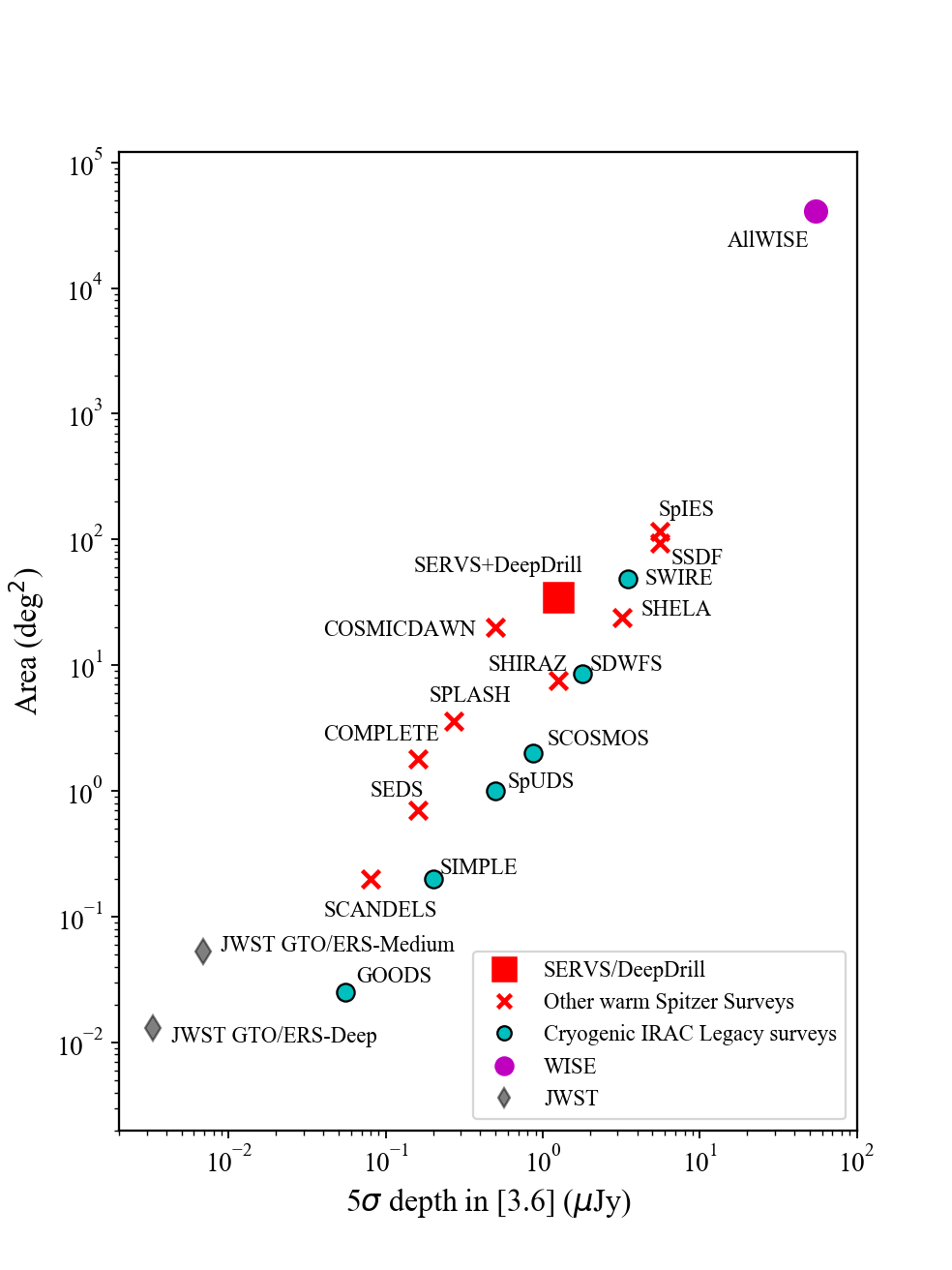}
    \caption{Depth versus area for extragalactic surveys at 3.6\,$\mu$m. Red crosses indicate surveys taken during the post-cryogenic phase of {\em Spitzer} as Exploration Science or Frontier Legacy surveys (surveys which incorporate previous efforts in the same fields have been combined). For comparison, we show surveys taken during the cryogenic mission of {\em Spitzer} as cyan circles, {\em WISE} mission in magenta, and two Guaranteed Time/Early Release (GTO/ERS) surveys planned for the {\em James Webb Space Telescope} in grey. References (top to bottom): AllWISE: \citet{2010AJ....140.1868W,2011ApJ...731...53M}; SpIES: \citet{2016ApJS..225....1T}; SSDF: \citet{2013ApJS..209...22A}; SWIRE: \citet{2003PASP..115..897L}; SERVS$+$DeepDrill: this paper; SHELA: \citet{2016ApJS..224...28P}; COSMICDAWN \citet{2016sptz.prop13058C}; SDWFS \citet{2009ApJ...701..428A}; SHIRAZ: Annunziatella et al.\ in preparation; SPLASH: \citet{2014ApJ...791L..25S}; SCOSMOS: \citet{2007ApJS..172...86S}; COMPLETE: \citet{2016sptz.prop13094L}; SpUDS \citet{2011AAS...21733551K}; SEDS: \citet{2013ApJ...769...80A}; SCANDELS \citet{2015ApJS..218...33A}; SIMPLE: \citet{2011ApJ...727....1D}; GOODS: \citet{2003mglh.conf..324D}; {\em JWST} ERS/GTO surveys: \citet{2019BAAS...51c..45R}.}
    
    \label{fig:area_depth}
\end{figure}

The observations described in this
paper supplement an earlier set of observations over smaller areas in these three fields by the {\em Spitzer} Extragalactic Representative Volume Survey \citep[SERVS;][]{2012PASP..124..714M}, for which images and catalogs are available from the Infrared Science Archive (IRSA) (a second data release of SERVS, including the data fusion of \citet{2015fers.confE..27V} and deeper {\em Spitzer} catalogs is planned). The DeepDrill images are of similar depth to those from SERVS (a $5\sigma$ depth of $\approx 2\,\mu$Jy in both bands), but cover more than twice the area ($\approx$ 27~deg$^2$ compared to 12~deg$^2$ in these fields covered by SERVS (see Table \ref{tab:observations}), though SERVS also includes a further 6~deg$^2$ in the Lockman and ELAIS-N1 fields). We also note that deeper warm {\em Spitzer} data in the ECDFS field were taken recently as part of the ``Cosmic Dawn Survey" (principal investigator P.\ Capak). Figure \ref{fig:area_depth} shows the SERVS-DeepDrill survey in the context of other surveys at $\approx$3.6\,$\mu$m.
The IRAC image and catalog data on the three DDFs described in this paper will be made available though the Infrared Science Archive (IRSA).

The scientific motivation for this survey closely followed that for SERVS, namely the study of galaxy evolution as a function of environment from $z\sim 5$ to the present, but with the additional feature of the deep and multi-epoch LSST DDF data. 
The DDFs are expected to be observed throughout the ten year duration of the LSST survey, with a cadence as frequent as once every two nights at times of year when the fields are available for observation \citep{2018arXiv181106542B,2018arXiv181200516S}. 
The time domain adds the ability to detect and obtain light curves of supernovae in distant galaxies \citep[of which we expect $\sim 10^4$ in the DDFs;][] {2009arXiv0912.0201L}, and to allow the study of AGN flares, tidal disruption events, and other variable phenomena. The {\em Spitzer} data provide information on the properties of host galaxies of supernovae  and other transients and help to identify and classify AGN, and, indeed, SERVS has already proved useful for these types of investigations
\citep{2014ApJ...787..138L, 2015A&A...579A.115F,2018MNRAS.478.2132C}. In conjunction with other data at optical and shorter near-infrared wavelengths, the {\em Spitzer} survey in these fields will enhance the study of the host galaxies of supernovae and AGN through improved estimates of stellar mass, star formation history and reddening \citep{2013MNRAS.435.1389P}. 

Medium-depth surveys with warm {\em Spitzer} covering areas $\sim 10-100$~deg$^2$ have proven very valuable for both studies of individual rare objects (with comoving densities $\sim 10^{-5}$ to $10^{-8}$~Mpc$^{-3}$), and 
statistical studies of populations including luminous AGN and quasars ($\sim 100$ deg$^{-2}$ in such surveys), galaxy clusters ($\sim 10$ deg$^{-2}$) and ultraluminous dusty star-forming galaxies ($\sim 1000$ deg$^{-2}$). Examples from SERVS include gravitational lenses, Lyman-$\alpha$ nebulae \citep{2018ApJ...854..151M,2019arXiv190708486M} and galaxy clusters at $z\sim 0.3-2$ \citep{2016A&A...592A.161N,2017MNRAS.465L.104N, 2017ApJ...843..126D,2018ApJ...866..136F,2019ApJ...880..119C, 2019ApJ...876...40P,2020MNRAS.493.5987O,2020arXiv200410757V}. 

SERVS has proven particularly valuable for the identification of the host galaxies of radio sources. These galaxies typically have high stellar masses, and are bright in the IRAC bands, with $>95$\% of faint radio sources identified in SERVS \citep[e.g.][]{2015AJ....150...87L, 2015MNRAS.453.4244W,2016MNRAS.463.2997M,2017MNRAS.468.1156O,2017MNRAS.470.4956S, 2018ApJ...856...67C, 2018MNRAS.481.4548P,2020MNRAS.491.1127O,2020MNRAS.497.5383I}. The small population of infrared-faint radio sources (IFRS) that are unidentified, or very faint, in the IRAC bands seem to represent a population of dust-reddened, high-$z$ radio-loud AGN \citep{2011ApJ...736...55N,2014A&A...567A.104H, 2016A&A...596A..80M}. 


Dusty star-forming galaxies detected in the mm/submm with positions from ALMA can often be identified with faint IRAC sources, allowing 
better understanding of their stellar masses and extinctions \citep{2014ApJ...788..125S, 2019ApJ...872..117G, 2019ApJ...871...85L, 2019ApJ...871..109P, 2020MNRAS.494.3828D,2020MNRAS.491.1127O,2020MNRAS.491.5911O}.
Other uses include obtaining constraints on stellar masses and ages of galaxies in overlapping deep spectroscopic surveys \citep{2017A&A...601A..95C,2017A&A...602A..35T,2019arXiv190301884K}, studying cosmic background radiation \citep{2016ApJ...832..104M} and exploiting fields suitable for deep multi-conjugate adaptive optics observations of distant galaxies \citep{2018ApJ...864....8L}.

\begin{figure*}
\centering
\includegraphics[scale=0.8]{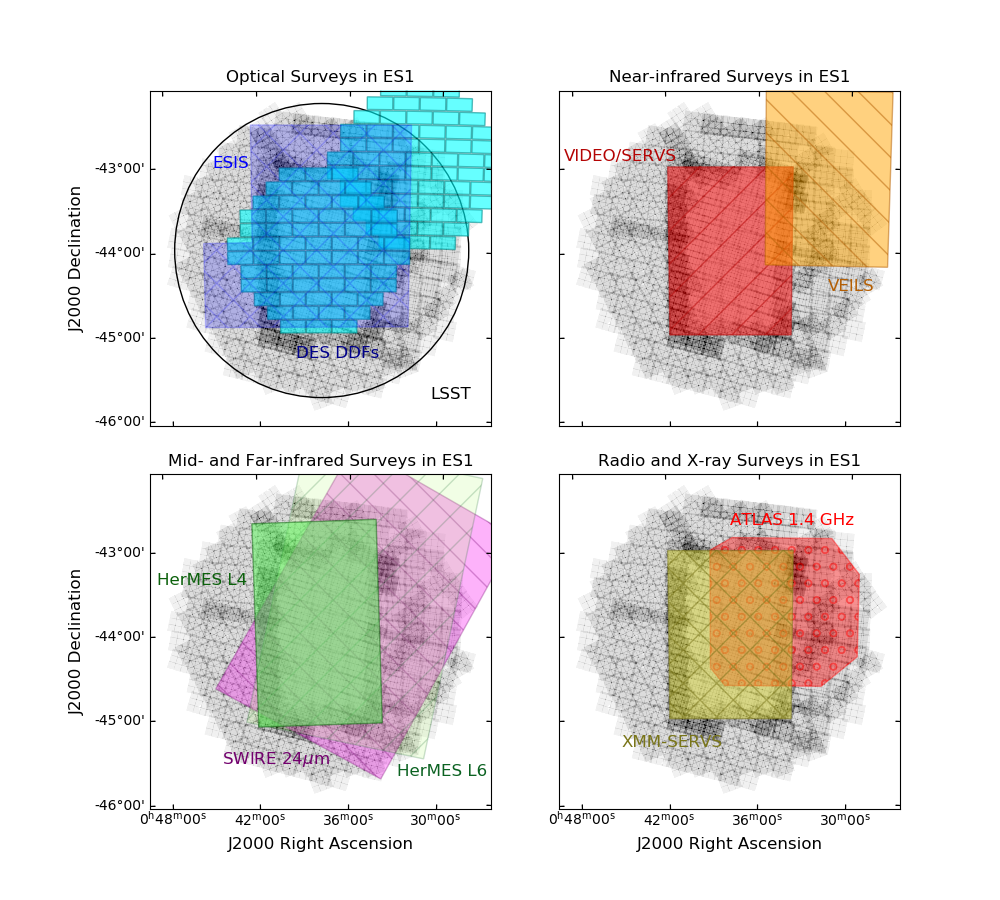}
\caption{The footprints of multi-wavelength surveys on the ELAIS-S1 field (see Table 2 for survey references), superimposed on a greyscale of the IRAC 4.5~$\mu$m coverage. {\em Upper left:} optical surveys (the DES DDFs are shown in cyan with rectangles indicating the individual chips, the ESIS survey in mauve with cross-hatches, and the LSST footprint is shown as a black circle), {\em upper right}, near-infrared surveys (VIDEO in red, left hatched and VEILS in orange, right-hatched), {\em lower left}, 24 $\mu$m coverage from SWIRE in magenta and far-infrared coverage in HerMES in green (the L4 data is deeper than the hatched L6 data), and {\em lower right} the X-ray {\em XMM}-SERVS coverage in yellow, cross-hatched and the ATLAS radio survey in red, with circular hatches.}
\label{fig:ES1}
\end{figure*}
\begin{figure*}
\includegraphics[scale=0.8]{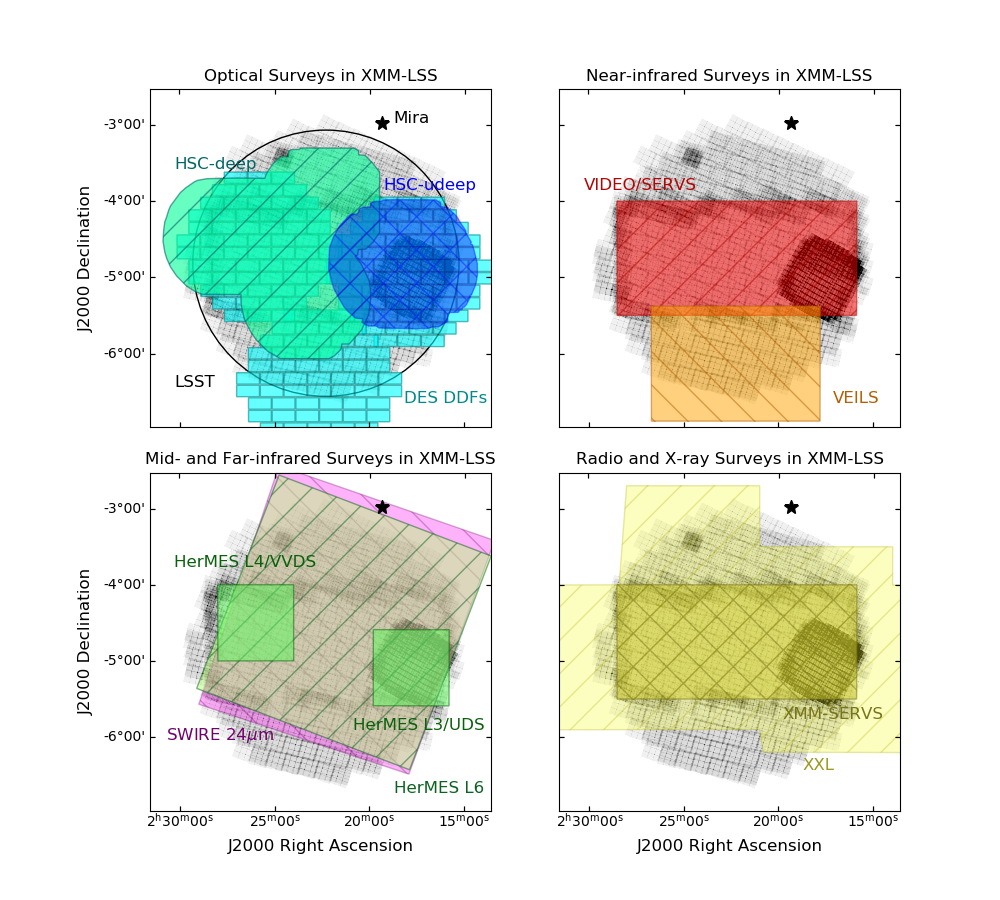}
\caption{The footprints of multi-wavelength surveys on the {\em XMM}-LSS field (see Table 2 for survey references), superimposed on a greyscale of the IRAC 4.5~$\mu$m coverage. {\em Upper left:} optical surveys (the LSST footprint is shown as a black circle, the DES DDFs are in cyan with rectangles indicating the individual chips, the HSC-deep survey is in green with left hatches, the HSC-ultradeep in blue with cross-hatching). {\em Upper right:} near-infrared surveys (VIDEO in red, left hatched and VEILS in orange, right-hatched), {\em lower left:} 24 $\mu$m coverage from SWIRE in magenta and far-infrared coverage from HerMES in green (the L3 data is the deepest, L4 less deep and L6 the shallowest). {\em Bottom right:} X-ray surveys - the XXL survey coverage is shown in light yellow, left-hatched and the {\em XMM}-SERVS survey in dark yellow, cross-hatched). The position of the infrared-bright, variable star Mira ($K\approx -2.2$) is indicated with the black star symbol.}
\label{fig:XMM}
\end{figure*}
\begin{figure*}
\includegraphics[scale=0.8]{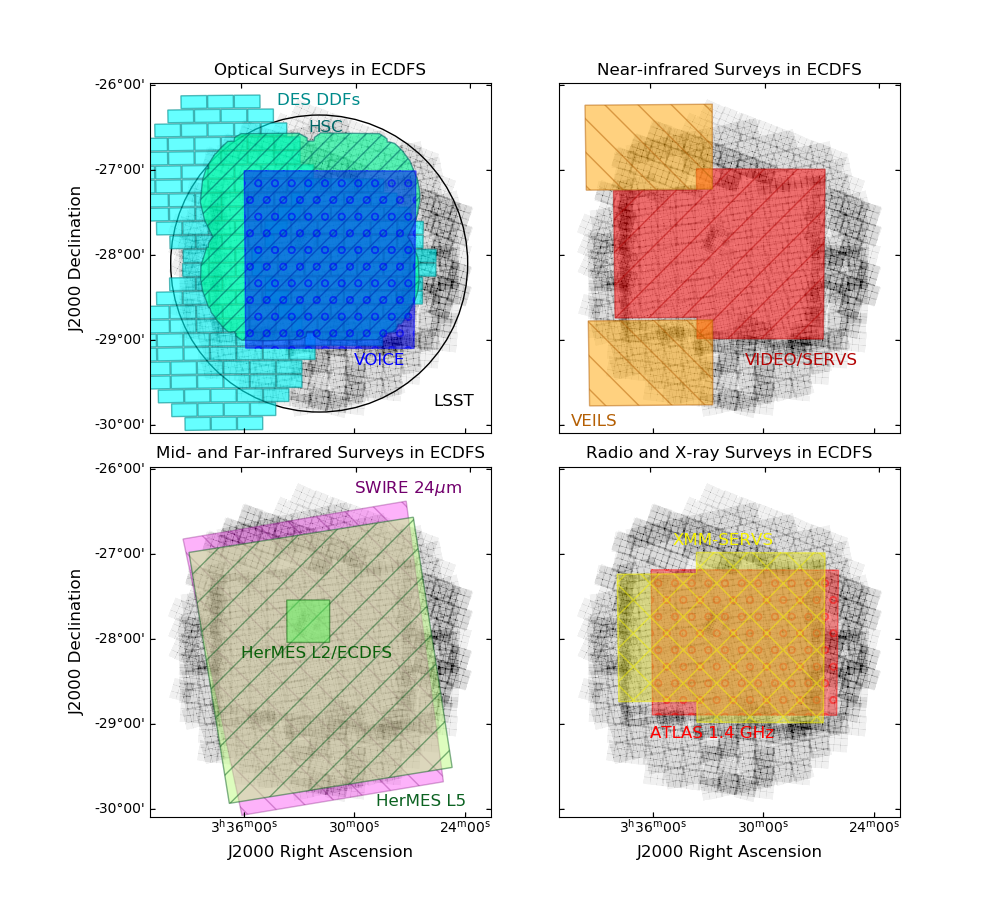}
\caption{The footprints of multi-wavelength surveys on the ECDFS field (see Table 2 for survey references), superimposed on a greyscale of the IRAC 4.5~$\mu$m coverage. {\em Upper left:} optical surveys (the LSST footprint is shown as a black circle, the DES DDFs are in cyan with rectangles indicating the individual chips, the HSC-deep survey in is green with left hatches, and the VOICE survey is in blue with circular hatches). {\em Upper right:} near-infrared surveys (VIDEO in red, left hatched and VEILS in orange, right-hatched), {\em lower left}, 24 $\mu$m coverage from SWIRE in magenta and far-infrared coverage in HerMES in green (the L2 data is deeper than the hatched L6 data), and {\em lower right} the X-ray {\em XMM}-SERVS coverage in yellow, cross-hatched and the ATLAS radio survey in red, with circular hatches.}
\label{fig:ECDFS}
\end{figure*}

The DDFs have garnered significant observational resources from other telescopes, from the radio and far infrared through to the X-ray. A list of the large area ($>1\;{\rm deg}^2$) surveys in the DDFs may be found in Table \ref{tab:surveys} \citep[see also Table 1 of][]{2018MNRAS.478.2132C}, and their coverages are illustrated in Figures \ref{fig:ES1}, \ref{fig:XMM} and \ref{fig:ECDFS}. In the X-ray, the {\em XMM}-SERVS survey \citep{2018MNRAS.478.2132C} is covering the original SERVS
areas in ES1, {\em XMM}-LSS and ECDFS. The optical data are less homogeneous, including data from Hyper Suprime-Cam (HSC) \citep{2018PASJ...70S...8A,Ni_2019}, the Dark Energy Survey \citep[DES;][]{2018ApJS..239...18A}, and the ESO ESIS and VOICE surveys
\citep{2006AA...451..881B,2016heas.confE..26V}, however, as all three fields will be targeted for deep LSST observations this is not a major concern. In the near-infrared, the VISTA VIDEO survey \citep{2013MNRAS.428.1281J} covers the whole SERVS area, and is supplemented by VEILS \citep{2017MNRAS.464.1693H} which covers the DES fields that are repeatedly observed to find supernovae and other time-domain phenomena (hereafter the DES DDFs). The fields are covered by the SWIRE survey in the mid-infrared \citep{2003PASP..115..897L}, and the HerMES survey in the 
far-infrared \citep{2012MNRAS.424.1614O}. In the radio, existing deep surveys from the ATCA (ATLAS) \citep{2015MNRAS.453.4020F}, GMRT \citep{2018AA...620A..14S} and LoFAR \citep{2019AA...622A...4H} cover a significant fraction of the fields. The MIGHTEE survey with MeerKAT, currently underway, will image the inner regions of all three fields even more deeply at 0.9--1.7~GHz \citep{2016mks..confE...6J}.

\citet{2015fers.confE..27V} combined SERVS data with catalogues of optical and near-infrared photometry that were available at the time in all five SERVS fields, and \citet{2019MNRAS.483.3168P} used these catalogues to derive photometric redshifts for $\approx 4$ million galaxies. Furthermore, the {\em Herschel} Extragalactic Legacy Project has incorporated SERVS data within their workflows to produce multi-wavelength catalogues and extract more accurate FIR/SMM fluxes
to study the dust properties of infrared galaxies over cosmic time
\citep{2016ASSP...42...71V,2017MNRAS.464..885H,2018A&A...620A..50M,2019MNRAS.490..634S}. For very challenging applications, such as identifying rare sources, and obtaining photometric redshifts accurate enough to study environments, more accurate photometry that allows for the difference between the relatively large {\em Spitzer} point spread function and overlapping ground-based surveys in the near-infrared or optical is needed. This more refined photometry requires the application of forced photometry techniques such as The {\sc tractor} \citep{2016ascl.soft04008L}, and has been successfully used in the {\em XMM}-LSS field \citep{2017ApJS..230....9N}, with the remaining SERVS/DeepDrill fields to follow. The improved photometry and photometric redshifts from it enable the accurate estimation of environmental parameters for galaxies out to at least $z\sim 1.5$ \citep{2020ApJ...889..185K}. The {\sc tractor} photometry in {\em XMM}-LSS was also used to obtain photometric redshifts for X-ray AGN in the {\em XMM}-SERVS survey \citep{2018MNRAS.478.2132C}.

This paper is structured as follows: Section \ref{sec:obs} describes the observations, Section \ref{sec:processing} the processing of the image data and tests to assess the quality of the astrometric and photometric calibration. Section \ref{sec:products} describes the image and catalogue data products to be included in the release. In Section \ref{sec:galaxies}, we present an overview of the galaxy population in DeepDrill, including colours and source counts, and also highlight sources with very red [3.6]$-$[4.5] colours found in the survey. Section \ref{sec:summary} contains a short summary.

\begin{table*}
\caption{{\em Spitzer}/IRAC DeepDrill Observations}
\begin{tabular}{lcccccc}
Field Name & DeepDrill Field Centre &SERVS Area$^*$ &DeepDrill & Total Area$^{\dag}$ 3.6\,$\mu$m & Total Area$^{\dag}$ 4.5\,$\mu$m & Total Area$^{\dag}$ Both\\
                    &    (J2000)              & (deg$^2$) & Observation dates&(deg$^2$)& (deg$^2$) & (deg$^2$) \\\hline\hline
 ES1  & 00:37:48 $-$44:01:30 & 3 & 2015-09-27 to 2016-10-24&9.2&9.0&8.6 \\
 {\em XMM}-LSS & 02:22:18 $-$04:49:00 & 4.5 &2015-10-21 to 2016-11-25 &9.2&9.4&8.9 \\
 ECDFS    & 03:31:55 $-$28:07:00 & 4.5& 2015-05-04 to  2016-12-26  &9.1&9.4&8.8\\\hline
\end{tabular}\label{tab:observations}
\begin{flushleft}
\noindent
$^{*}$ The SERVS field centres differ slightly from the DeepDrill ones, but the SERVS fields are entirely encompassed by the DeepDrill survey.

\noindent
$^{\dag}$ Total areas are those covered by the SERVS and DeepDrill data combined.
\end{flushleft}
\end{table*}

\begin{table*}
\caption{Other surveys ($>1$deg$^{2}$) in the DeepDrill fields.}
\scriptsize{
\begin{tabular}{llllllr}
Survey & Field(s) & Bands/Wavelengths/Energies & Overlap Area & Depth$^{\ddag}$ & Reference\\ 
             &           &                   & (deg$^2$)   &                 & \\\hline\hline
XMM-SERVS & ECDFS, {\em XMM}-LSS, ES1 & 0.5-10~keV & 13 & $1.7 \times 10^{-15}{\rm erg\; cm^{-2} s^{-1}}$ (0.5-2 keV)&\citet{2018MNRAS.478.2132C} \\
XXL     & {\em XMM}-LSS & 0.5-10~keV &  8 & $5 \times 10^{-15}{\rm erg\; cm^{-2} s^{-1}}$ (0.5-2 keV)& \citet{2016AA...592A...1P} \\
DEVILS & ECDFS, {\em XMM}-LSS & 3750-8850\AA & 4 & Spectroscopic, $Y<21.2$ & \citet{2018MNRAS.480..768D}\\
ESIS   & ES1   & $B,~V,~R$       & 4.5& Vega magnitude $\approx 25$&\citet{2006AA...451..881B}\\
VOICE & ECDFS &  $u,~g,~r,~i$ & 8 & $AB$ magnitude $\approx 26$ & \citet{2016heas.confE..26V}\\
DES (DR1) & ES1, {\em XMM}-LSS, ECDFS & $g,~r,~i,~z$ & 28 & $AB=25.1,\, 24.8,\, 24.2,\, 23.4,\, 22.2^{\ddag\ddag}$ & \citet{2018ApJS..239...18A}\\
HSC (DR1) & {\em XMM}-LSS & $g,~r,~i,~z,~y$& $\approx 6$  & $i_{AB}\approx 26.5-27.0^{\S}$ & \citet{2018PASJ...70S...8A}\\ 
HSC (Ni et al.) & ECDFS & $g,~r,~i,~z$ & 5.7 & $AB\approx 25.9,~25.6,~25.8,~25.2$ &\citet{Ni_2019}\\
SWIRE &ECDFS, {\em XMM}-LSS, ES1&  3.6, 4.5, 5.8, 8.0, 24, 60, 160 ~$\mu$m & 27 &  various depths       &    \citet{2003PASP..115..897L}       \\
OzDES & ES1, {\em XMM}-LSS, ECDFS & Spectroscopic; 370-880 nm&17 & $r_{AB}\approx 23$&\citet{2017MNRAS.472..273C}\\
PFS & XMM-LSS & Spectroscopic; 380-1260 nm& 6 & $J_{AB} \approx 23.4$ & \citet{2014PASJ...66R...1T}\\
VIDEO &ECDFS, {\em XMM}-LSS, ES1 & $(Z)^{*}, (Y)^{*}, J, H, K_{s}$ &   13 & (25.7), (24.6), 24.5, 24.0, 23.5& \citet{2013MNRAS.428.1281J} \\
VEILS & ECDFS, {\em XMM}-LSS, ES1 & $J,\, K_{s}$ & $\approx 6$ & 25.5, 24.5 & \citet{2017MNRAS.464.1693H}\\
HerMES &ECDFS, {\em XMM}-LSS, ES1&  250, 350, 500~$\mu$m &    27    &   $\sim 25$mJy${\S}$ &\citet{2012MNRAS.424.1614O}\\
ATLAS & ECDFS, ES1 & 1.4~GHz & 6.3 & 14-17$\mu$Jy & \citet{2015MNRAS.453.4020F}\\
GMRT  & {\em XMM}-LSS & 610~MHz & 8 & $\approx 1\;$mJy & \citet{2018AA...620A..14S}\\
LoFAR & {\em XMM}-LSS & 120--168~MHz & 9 & 1.4~mJy & \citet{2019AA...622A...4H}\\
MIGHTEE& ECDFS, {\em XMM}-LSS, ES1 & 900-1670~GHz$^{**}$ & 16.6 & 2$\mu$Jy & \citet{2016mks..confE...6J}\\\hline
\end{tabular}
\label{tab:surveys}
\begin{flushleft}
\noindent
$^{\ddag}$ Typical source detection limit ($\approx 5\sigma$).

\noindent
$^{*}$ ECDFS was only observed in $J, H$ and $K_{s}$.

\noindent
$^{**}$ A smaller area survey (4 deg$^2$) will also be carried out at 2-4~GHz in ECDFS.

\noindent
$^{\S}$ The {\em XMM}-LSS field of the HSC survey contains one ultradeep pointing and three deep ones, so the depth varies with position.

\noindent
$^{\ddag\ddag}$ 10$\sigma$ magnitude limits from \citet{2018ApJS..239...18A} $+0.75$ to convert to $5\sigma$; note that there is significant overlap between DeepDrill and the DES Deep Drilling fields (see Figures \ref{fig:ES1}-\ref{fig:ECDFS}, which, when the data are co-added, will be significantly deeper than the main survey).

\noindent
$^{\S}$ \citet{2017MNRAS.464..885H}
\end{flushleft}
}
\end{table*}

\section{Observations}\label{sec:obs}

We were awarded time to perform a survey of three of the four LSST DDFs that have been defined at the time of writing: \footnote{\url{https://www.lsst.org/scientists/survey-design/ddf}} the ELAIS-S1 field (ES1), the {\em XMM}-Large-Scale Structure Survey field ({\em XMM}-LSS) and the Extended {\em Chandra} Deep Field-South field (ECDFS). The fourth DDF identified by LSST, the COSMOS field, has deep coverage (to  $5\sigma$ depth of $\approx 0.3\,\mu$Jy in both bands) in the inner 2~deg$^2$ from several {\em Spitzer} surveys \citep{2007ApJS..172...86S, 2014ApJ...791L..25S, 2018ApJS..237...39A}. There is also a wider survey (SHIRAZ; Annunziatella et al. 2020 in preparation), to a similar depth as SERVS/DeepDrill that covers an additional $\approx$ 2~deg$^2$ outside of the central area to overlap with the Hyper Suprime-Cam Deep Survey \citep{2018PASJ...70S...8A}. 

\begin{figure}
    \centering
    \includegraphics[scale=0.9]{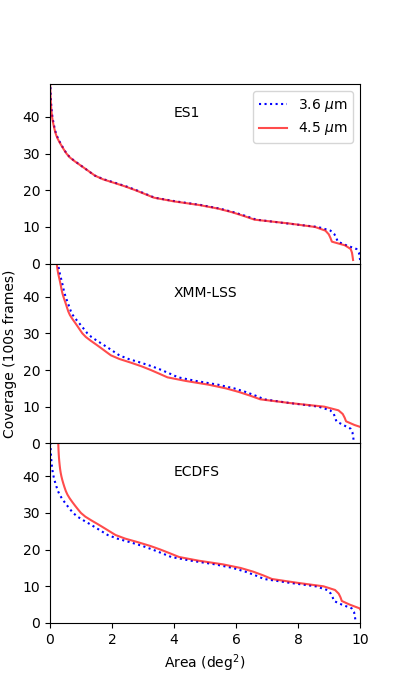}
    \caption{Area of each field as a function of coverage in units of IRAC image frames of 100s duration. The distribution for the 3.6 $\mu$m band is shown as the blue dotted line, and that for the 4.5 $\mu$m band as the continuous red line. The small areas of very high ($>50$) frame coverage in XMM-LSS and ECDFS result from the inclusion of data from earlier deep surveys in these fields, as detailed in Section \ref{sec:processing}.}
    \label{fig:coverage}
\end{figure}

The central areas of all three of our fields were observed as part of SERVS \citep{2012PASP..124..714M} during the early months of the post-cryogenic {\em Spitzer} mission
(2009-07-28 to 2011-03-06). The DeepDrill Survey (Program ID 11086, P.I.\ Lacy) was observed between 2015-05-04 and 2016-12-26 (Table \ref{tab:observations}). The DeepDrill observations followed the SERVS Astronomical Observation Request (AOR) construction, with each AOR making up a square tile of nine pointings, each pointing consisting of six repeats of 100s frames dithered using the IRAC small cycling dither pattern.\footnote{See the IRAC Instrument Handbook \url{https://irsa.ipac.caltech.edu/data/SPITZER/docs/irac/iracinstrumenthandbook/}} The use of Fowler sampling in the IRAC detectors \citep{2004ApJS..154...39F} means that the 100s frame time corresponds to a little less than 100s integration on sky: 93.6s at 3.6 $\mu$m and 96.8s at 4.5 $\mu$m. The fields were imaged in two epochs to facilitate rejection of asteroids, with a targeted depth of 12 frames. Due to scheduling constraints, the time separation of the two epochs was non-uniform, ranging from a few weeks to $\sim 1$ year. The  spatial coverage is also non-uniform. Areas around the edges of the SERVS fields in particular received additional coverage, and some outlying regions did not receive the full coverage. Figure \ref{fig:coverage} shows the distribution of coverage in each field. The area in each band with a coverage of 9 or more 100-second frames (i.e.\ with $>$87\% of the sensitivity of the nominal coverage of 12 frames), and the area with coverage of 9 or more in both bands at the same position are listed in Table \ref{tab:observations} (the area with both bands is slightly smaller as the two IRAC detectors are offset). We encourage users with a need for uniformity in depth to make use of the supplied coverage maps. 

The survey was designed such that source confusion only becomes significant near the nominal flux density limit of the survey, where there are about 30 beams per source, the typical value at which source confusion becomes significant \citep{2012ApJ...758...23C}. In Appendix \ref{sec:confusion}, we 
show how the confusion noise is expected to vary with depth of coverage in the survey, including both confusion from randomly-distributed sources and an additional term due to galaxy clustering. 
To more accurately extract faint source parameters from the deepest parts of the survey we recommend PRF fitting of sources and their near-neighbors, which can be further improved by using a prior from a higher resolution survey of similar or greater depth.

\section{Data analysis}\label{sec:processing}

\subsection{Image processing}
Data processing of the DeepDrill data was similar to that carried out for the SERVS dataset \citep{2012PASP..124..714M}, using a data cleaning pipeline derived from processing SWIRE and COSMOS data \citep{2003PASP..115..897L, 2007ApJS..172...86S}. The processing began with the Corrected Basic Data (CBCD) frames from the {\em Spitzer} Science Center \citep[basic calibrated data frames with corrections for common artifacts, see][]{2016SPIE.9904E..5ZL}. A refined dark frame for each AOR was constructed after identifying and masking astronomical sources in the data and subtracted from each individual frame in the AOR. Hot pixels in the 4.5$\mu$m data were masked, and the column pulldown correction provided in the CBCD was improved \citep[see][]{2012SPIE.8442E..38L}. The images were then rectified to a common background level corresponding to the mean background during the observations. Further corrections were made for latent images on a frame-by-frame basis, as these are particularly prevalent in warm mission data. The data were then mosaicked using {\sc mopex} \citep{2006SPIE.6065..330M} \citep[see table 3 of][for the parameters used]{2012PASP..124..714M}.

A pointing issue was found and corrected in the ES1 field, where the pointing refinement task in the {\em Spitzer} data processing pipeline failed for four of the AORs in the South of the field. We also found that we needed to correct the photometric calibration of the SERVS data in ES1, which was taken early in the post-cryogenic mission, while the instrument performance was still being characterized and before the final array temperatures had been set. This was done by comparing the fluxes of sources in the overlap between the SERVS and DeepDrill datasets, and applying the measured offsets to the SERVS data (1.04 at [3.6] and 0.98 at [4.5]). The ES1 data are thus all on the final warm mission calibration. Following \citet{2012PASP..124..714M}, the $\approx 1$ deg$^2$ in the Southwestern part of the {\em XMM}-LSS field that was taken during the cryogenic mission as part of the SpUDS programme \citep{2011AAS...21733551K} was included in the final images. The calibrations of these data are the same as those of the DeepDrill data to within 1\%, but the SpUDS data are significantly deeper. Similarly, the central $\approx 0.5$ deg$^{2}$ of the ECDFS field contains data from the much deeper cryogenic SIMPLE \citep{2011ApJ...727....1D} and GOODS \citep{2003mglh.conf..324D} programmes. However, in this case we used only a depth of $\approx 12$ frames per sky position to obtain an approximately uniform depth for the survey of that field. 


\subsection{Astrometric accuracy}

We matched the DeepDrill catalogs to {\em Gaia} Data Release 2 \citep[DR2;][]{2018A&A...616A...2L}. 
The IRAC pointing is calibrated using 2MASS \citep{2006AJ....131.1163S}, and the pointing refinement pipeline now includes proper motion information to provide positions for epoch J2000 \citep{2016SPIE.9904E..5ZL}. We therefore used the proper motion information in {\em Gaia} to derive positions appropriate for the year 2000 to match to the {\em Spitzer} positions. We matched the dual-band DeepDrill catalogues to Gaia using a 1\farcs0 match radius. 3\% of sources in the DeepDrill survey have counterparts in Gaia DR2. The results are shown in Table \ref{tab:positions}, where we list the mean systematic offset between {\em Spitzer} and {\em Gaia} DR2 positions $\Delta$(R.A.) and $\Delta$(Decl.), along with the scatter $\sigma$(R.A.) and $\sigma$(Decl.), representing the positional accuracy of a typical source in the survey. This scatter is independent of source flux, and is thus probably dominated by the scatter in the pointing refinements of individual frames, which is $\approx$0\farcs3 on an individual CBCD frame \citep{2016SPIE.9904E..5ZL}. All systematic offsets when averaged over a mosaic are $<0.1$~arcsec. 

\begin{table}
\caption{IRAC source matches in {\em Gaia} DR2: number of matches, mean position offsets and scatter.}
\begin{tabular}{lccccc}\label{tab:positions}
Field& Matches & $\Delta$(R.A.)  & $\Delta$(Dec.) & $\sigma$(R.A.) & $\sigma$(Dec.)\\
           & & (arcsec) & (arcsec) & (arcsec) & (arcsec) \\\hline\hline
 ES1  & 24375& 0.07 & 0.00 & 0.24 & 0.22 \\
 {\em XMM}-LSS  &19712 &$-$0.04& 0.02 & 0.23 & 0.21 \\
 ECDFS    &23946 &$-$0.02 & 0.02& 0.20 & 0.21   \\\hline
\end{tabular}
\end{table}

\subsection{Photometric accuracy}

The photometric calibration history of IRAC is given in \citet{2012SPIE.8442E..1ZC}. The post-cryogenic mission calibration factors were obtained by matching the fluxes of the standard stars to the
cryogenic observations, and have an absolute accuracy $\approx 3$\%. Some scatter is to be expected, as the $\approx 1\farcs 8$ IRAC PSF is undersampled by the 1\farcs2 pixels, and the sensitivity across each pixel varies as a function of position within the pixel. As most of the sources in DeepDrill are slightly extended, and because any given point in the sky is covered by a large number of observations, this intrapixel sensitivity variation is assumed to average out. Aperture corrections were applied as described in \citet{2012PASP..124..714M}, using the values in Table 2 of that paper.

As noted in Section \ref{sec:processing}, some SERVS fields (including ES1) were taken early in the {\em Spitzer} warm mission, before the calibration factors were finalized. In SERVS, this issue was dealt with by forcing the flux densities to the same scale as SWIRE. The SWIRE flux densities were based on an early calibration of IRAC, and thus there are significant differences between the calibration of the DeepDrill data (which are based on the final {\em Spitzer} post-cryogenic mission calibration) and the original 
SERVS data in the same fields. Table \ref{tab:photo} gives the ratio of the calibration factors derived from comparing the DeepDrill to the SERVS Aperture 2 (3\farcs9 diameter) flux densities for sources $>10\,\mu$Jy in the same field. The more accurate DeepDrill calibration is preferred.

\begin{table}
    \caption{Differences in the photometic calibration between SERVS and DeepDrill (the DeepDrill calibration is preferred.)}
    \begin{tabular}{l|cc}
Field        &  \multicolumn{2}{c}{DeepDrill/SERVS flux ratio}\\
             &   3.6 $\mu$m & 4.5 $\mu$m \\\hline\hline
ES1     &   0.95$\pm 0.01$& 0.96$\pm 0.01$\\
{\em XMM}-LSS      &   0.96$\pm 0.01$ & 0.98$\pm 0.01$\\
ECDFS         & 0.95$\pm 0.01$ & 0.97$\pm 0.01$\\\hline
   \end{tabular}
    \label{tab:photo}
\end{table}

\section{Data products}\label{sec:products}

This section briefly describes the data products from DeepDrill that are available in the data release. These consist of images and two sets of catalogues, single-band catalogues and dual-band catalogues.


\subsection{Images}

Images were made at a pixel scale of 0\farcs60 per pixel (oversampling the PSF width of 1\farcs8), and are calibrated in MJy\;sr$^{-1}$. This results in a conversion factor from pixel values to $\mu$Jy of 8.46. Coverage images were made, in units of 100-second IRAC frames, along with uncertainty images. Finally, mask images, showing the location of bright stars in the fields are included, made using the methods described in \citet{2012PASP..124..714M}.

\subsection{Catalogues}


Single-band (3.6$\;\mu$m and 4.5$\;\mu$m) catalogues were produced using SEx{\sc tractor} \citep{1996A&AS..117..393B}. Parameters used are shown in Table \ref{tab:se}, note some of these differ from \citet{2012PASP..124..714M}, principally to improve background filtering: the background mesh size was changed from 32 to 16 pixels, and the filtersize from five to three. These changes improved the background estimates in regions of scattered light around bright objects. The default convolution filter (2-pixel [1\farcs2] FWHM) was used for source detection, along with a weight map (the depth of coverage map was used). Photometric apertures labelled 1--5 corresponded to the SWIRE \citep{2003PASP..115..897L} standard apertures with radii 1\farcs4, 1\farcs9, 2\farcs9, 4\farcs1 and 5\farcs8, respectively, with 
aperture corrections applied per \citet{2012PASP..124..714M}. Uncertainties in the flux densities are from SEx{\sc tractor}, adjusted to allow for the effects of pixel resampling and detector gain. An additional 3\% error is added in quadrature to account for the systematic error in the IRAC flux density scale.
The raw output from SEx{\sc tractor} was filtered to output sources with a signal-to-noise ratio $>5$ in the SWIRE aperture 2 (3\farcs9 diameter). The flag column in the catalogue is a bitwise flag that takes the first and second bit of the SEx{\sc tractor} flag (bit 1: photometry may be affected by neighbours or bad pixels, and bit 2: the source was blended with a neighboring object), and adds a further flag bit (3) to indicate that the source is either a bright ($K<12$) star, or within the region affected by the halo of a bright star according to the rules in \citet{2012PASP..124..714M}. The catalog columns are listed in Table \ref{tab:1chan}. The star masks used to create this flag are included in the data delivery to the Infrared Science Archive (IRSA). The 3.6$\;\mu$m and 4.5$\;\mu$m catalogues contain 2.7 and 2.5 million sources, respectively, summed over all three fields.

\begin{table}
\caption{Source extraction parameters used in SEx{\sc tractor}}
\label{tab:se}
\begin{tabular}{lc}
Parameter & Value\\\hline\hline
DETECT\_MINAREA & 5\\
DETECT\_THRESH  & 1.0\\
DETECT\_MINCONT & 0.0005\\
ANALYSIS\_THRESH & 0.4\\
BACK\_SIZE & 16\\
BACK\_FILTERSIZE & 3\\
BACKPHOTO\_TYPE & LOCAL\\\hline

\end{tabular}

\end{table}

\begin{table*}
    \caption{Single band catalogue columns}
    \centering
    \begin{tabular}{lll}
     Column(s)    & Description & Units \\ \hline\hline
      1   & Name (J2000 coordinates prefixed by DD1)& -\\
      2   & Right Ascension (ICRS)& Degrees\\
      3   & Declination (ICRS)& Degrees\\
      4-8 & Aperture corrected flux densities in apertures 1-5& $\mu$Jy\\
      9   & Isophotal flux density& $\mu$Jy\\
      10  & SEx{\sc tractor} Auto flux density& $\mu$Jy\\
      11-17& Uncertainties on columns 4-10& $\mu$Jy\\
      18 & Kron radius & 0\farcs6 pixels\\
      19 & Signal-to-noise ratio& -\\
      20 & Local RMS noise& $\mu$Jy\\
      21 & Coverage & 100s frames\\
      22 & Flag (see text)& -\\\hline
    \end{tabular}
    \label{tab:1chan}
\end{table*}
\begin{table*}
    \caption{Dual band catalogue columns}
    \centering
    \begin{tabular}{lll}
     Column(s)    & Description & Units\\ \hline\hline
      1   & Name (J2000 coordinates prefixed by DD1)& -\\
      2   & Right Ascension (ICRS)& Degrees\\
      3   & Declination (ICRS)& Degrees\\
      4-8 & Aperture corrected flux densities at 3.6$\mu$m in apertures 1-5& $\mu$Jy\\
      9-13 & Aperture corrected flux densities at 4.5$\mu$m in apertures 1-5& $\mu$Jy\\
      14-23& Uncertainties in columns 4-13& $\mu$Jy\\
      24-25  & Isophotal flux densities at 3.6 and 4.5$\mu$m& $\mu$Jy\\
      26-27  & Uncertainties in columns 24-25& $\mu$Jy\\
      28-29  & SEx{\sc tractor} Auto flux densities at 3.6 and 4.5$\mu$m& $\mu$Jy\\
      30-31  & Uncertainties in columns 28-29& $\mu$Jy\\
      32-33& Kron radii at 3.6 and 4.5$\mu$m& 0\farcs6 pixels\\
      34-35 & Signal-to-noise ratios at 3.6 and 4.5$\mu$m&-\\
      36-37 & Coverage & 100s frames\\
      38-39 & Flags at 3.6 and 4.5$\mu$m (see text)& -\\\hline
    \end{tabular}
    \label{tab:dual}
\end{table*}


Dual-band catalogues were created by matching the two single-band catalogues produced by SEx{\sc tractor} with a 0\farcs6 matching radius (before applying the $5\sigma$ cut) and then applying a 3$\sigma$ cut for the signal-to-noise ratio of the detection in a 3\farcs9 diameter at both 3.6\;$\mu$m and 4.5$\mu$m. (We considered using the dual-image capability of SE{\sc xtractor}, but we found that the approach of performing two independent source detection rounds on the individual channels and then merging the results in catalogue space resulted in a more reliable catalogue.) There will thus be objects present in the merged catalogue that are not present in the single-band catalogues (and vice-versa). The 3.6\;$\mu$m positions are given in the catalog as these correspond to the smallest PSF. Columns are listed in Table \ref{tab:dual}. The dual-band catalogues in each field contain approximately 800,000 sources, giving a total of 2.4 million sources.

Multiwavelength catalogues in the center 3--5 deg$^2$ of each of the DeepDrill fields are in the process of construction \citep[Nyland et al., in preparation, ][]{2017ApJS..230....9N}. These employ forced photometry with the {\sc tractor} \citep{2016ascl.soft04008L}, using the ground-based near-infrared VIDEO data as a prior to overcome source blending issues. Recovered IRAC magnitudes from this technique are a good match to the aperture magnitudes in SERVS/DeepDrill \citep{2017ApJS..230....9N}. These catalogues are not part of the initial DeepDrill data release, but will form part of a subsequent data release.



\begin{figure*}
    \centering
    \includegraphics[scale=0.95]{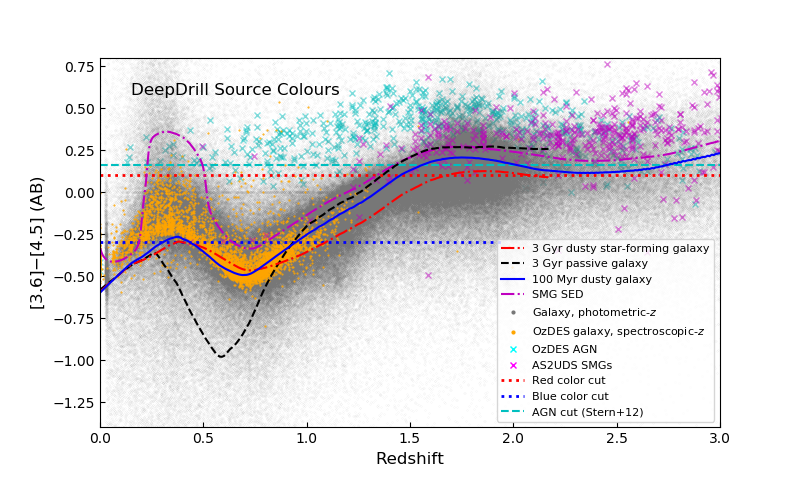}
    \caption{[3.6]$-$[4.5] colours of galaxies as a function of redshift \citep[cf.][]{2008ApJ...676..206P}. The grey dots show the galaxies with photometric redshifts in the {\em XMM}-LSS field \citep{2020ApJ...889..185K}, spectroscopic galaxy targets from the OzDES survey \citep{2017MNRAS.472..273C} (OzDES LRG, ClusterGalaxy, RedMaGiC, BrightGalaxy, Emission Line Galaxy and Photo-z types) are shown as orange dots, AGN\_monitoring targets from OzDES are shown as cyan crosses, and 
    A2SUDS submm galaxies with photometric redshifts as magenta crosses. The lines indicate various color cuts; the ``Blue cut'' (blue dotted line) is the line below which the blue population of Figure \ref{fig:counts} (d) is selected; ``red cut'' (red dotted line) the line above which the red population of Figure \ref{fig:counts} (d) is selected. We also include the [3.5]$-$[4.6] color cut of \citet{2012ApJ...753...30S} (``AGN cut''; cyan dashed line), above which candidate AGN dominate in the much shallower WISE survey. Three models based on \citet{2005MNRAS.362..799M} evolutionary synthesis models \citep[see][]{2019MNRAS.483.3060G} are shown, a 3 Gyr old dusty star-forming galaxy, a 3 Gyr passive galaxy and a 100 Myr galaxy. (Note that the 3 Gyr models do not extend beyond $z=2.17$, where the age of the Universe is 3 Gyr.) We also plot the composite SED of the A2SUDS submm galaxies from \citet{2020MNRAS.494.3828D}.}
    \label{fig:col_z}
\end{figure*}

\section{Analysis and results}\label{sec:galaxies}

\subsection{[3.6]--[4.5] colour as a function of redshift}\label{sec:color_redshift}

To show the variation of galaxy colour with redshift in DeepDrill we use the photometric redshifts in {\em XMM}-LSS from \citet{2020ApJ...889..185K}. These use {\sc tractor}-based forced photometry \citep[Nyland et al.\ 2020, in preparation, see also][]{2017ApJS..230....9N,2018ApJ...856...67C} to obtain redshift estimates for 690,000 sources in the overlap between SERVS and DeepDrill. These photometric redshifts have an uncertainty in $z/(1+z)\approx 0.03$ and outlier fraction 1.5\% in the redshift range $0<z<1.5$. To supplement these photometric redshifts with a smaller, but more accurate, sample of spectroscopic redshifts we matched to the OzDES Data Release 1 \citep{2017MNRAS.472..273C}, finding 9623 matches within 1$^{''}$ of the DeepDrill positions (corresponding to $\approx 0.4$\% of the galaxies in DeepDrill). The OzDES AGN were specially targeted for monitoring programmes. 

In the absence of photometric redshifts (still the case over much of the DeepDrill survey), the [3.6]$-$[4.5] colour can be used as a crude redshift indicator \citep{2008ApJ...676..206P}.
Figure \ref{fig:col_z} shows this  colour as a function of redshift.
The models \citep[based on][]{2005MNRAS.362..799M} show that the dip in the [3.6]$-$[4.5] colour at $z\approx 0.6$--$0.8$ is a strong function of the shape of the spectral energy distribution in the near-infrared, which is itself a strong function of the age and nature of the stellar population. Nevertheless, the spectroscopic and photometric redshifts show that most galaxies follow a fairly tight trend in [3.6]$-$[4.5] colour as a function of redshift. Galaxies with blue colours (bluer than the blue dotted line at [3.6]$-$[4.5]$=-0.3$ in Figure \ref{fig:col_z}) are at $z\approx 0.5$--$0.9$, and galaxies with red colours (redder than the red dotted line corresponding to [3.6]$-$[4.5]$=$0.1) are at $z\stackrel{>}{_{\sim}}1.3$.
Also shown in Figure \ref{fig:col_z} is the boundary color commonly used to select AGN candidates from WISE data \citep{2012ApJ...753...30S} :  [3.5]$-$[4.6]$>$0.8 in the Vega system, corresponding to [3.6]$-$[4.5]$>$0.16 in AB. We also plot the colours and photometric redshifts of submm-selected galaxies from the A2SUDS sample of \citet{2020MNRAS.494.3828D} and the locus of the [3.6]$-$[4.5] colour from their composite SED. The very red colour of the A2SUDS SED in the range $0.3\stackrel{<}{_{\sim}}z\stackrel{<}{_{\sim}}0.5$ is due to the strong 3.3$\mu$m Polycyclic Aromatic Hydrocarbon (PAH) feature passing through the [4.5] band.

\begin{figure*}
    \centering
    \includegraphics[scale=1.0]{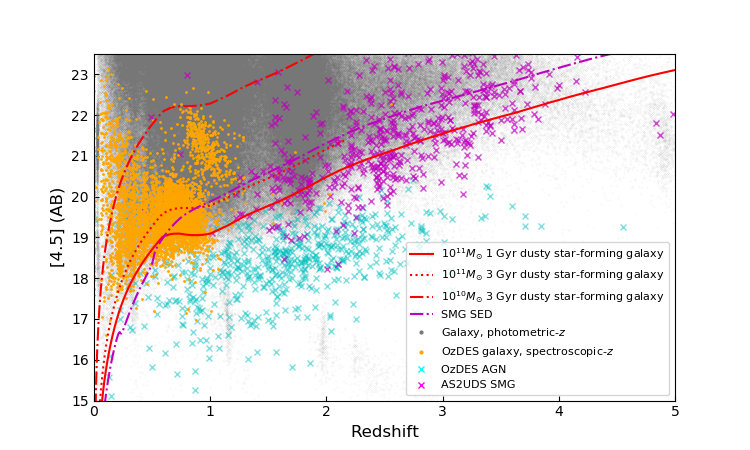}
    \caption{Magnitude versus redshift in [4.5] (symbols as for Figure \ref{fig:col_z}). Also plotted are two instances of the 3 Gyr old dusty star-forming galaxy model in Figure \ref{fig:col_z} with stellar masses of 10$^{10} M_{\odot}$ and
    10$^{11} M_{\odot}$, and a 1 Gyr dusty galaxy model with 10$^{11} M_{\odot}$ from the same library, along with the composite submm galaxy SED from A2SUDS.}
    \label{fig:hubble}
\end{figure*}

In Figure \ref{fig:hubble} we plot the [4.5] magnitude against redshift for the sample matched to OzDES, the A2SUDS sample and the galaxies with photometric redshifts (note that the ``banding'' in the distribution of photometric redshifts is an artifact of the photometric redshift algorithm). Most of the galaxies in the survey have stellar masses between $10^{10}$ and $10^{11} M_{\odot}$, and galaxies with masses of 10$^{11} M_{\odot}$ can 
be detected out to $z\approx 5$.
As expected, the AGN are red and bright, though it should be noted that around the AllWISE limit of $AB\, \approx \,19.5$ the counts of red galaxies
begin to rapidly exceed those of AGN, making selection of AGN based on [3.6]$-$[4.5] colour alone highly contaminated at faint magnitudes.

\subsection{Source counts}

\begin{figure*}
\includegraphics[scale=1.0]{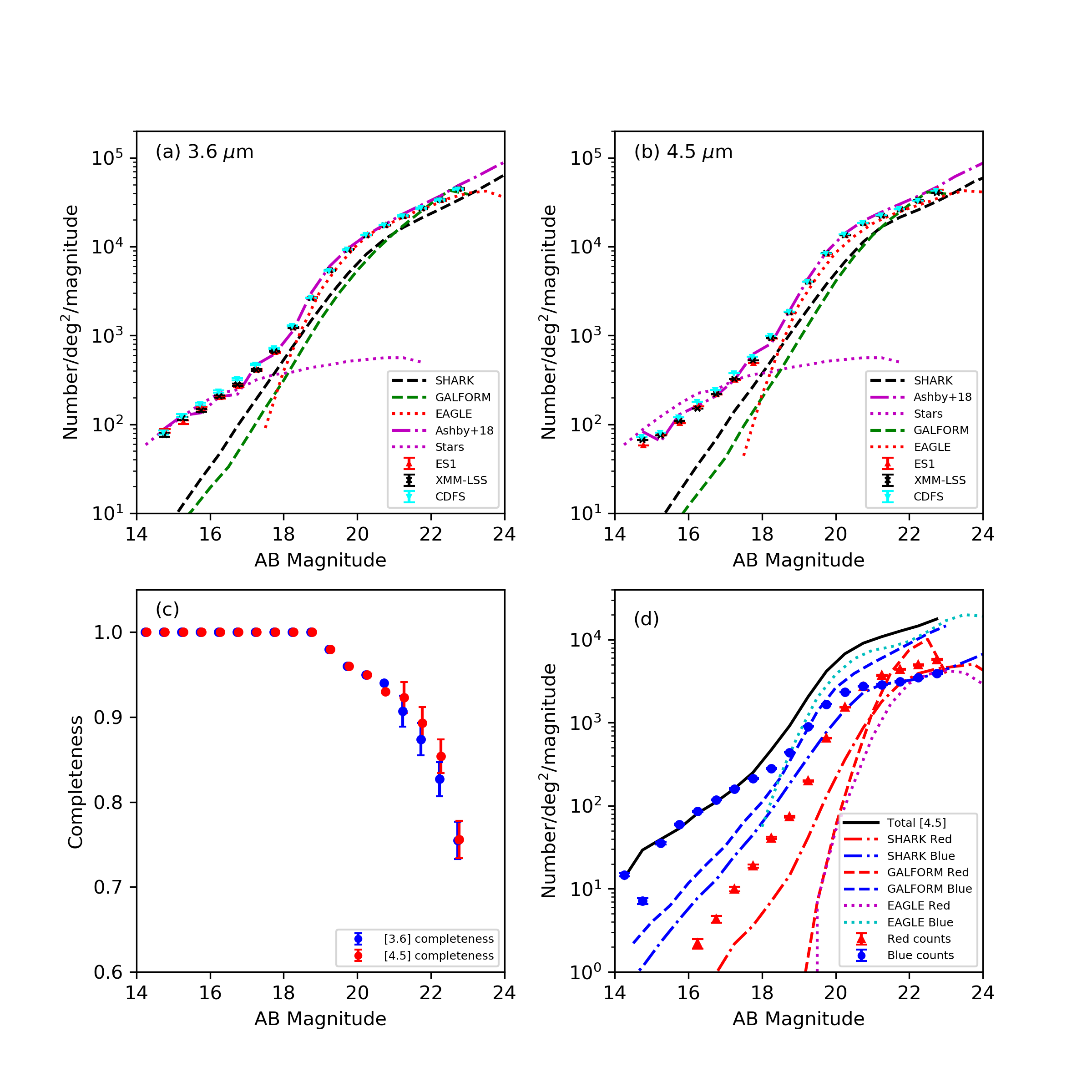}
\caption{DeepDrill source counts. (a) the 3.6~$\mu$m counts: the three sets of points show the completeness-corrected counts from each DeepDrill field. The dot-dashed magenta line shows the mean counts from the S-CANDELS fields \citep{2015ApJS..218...33A}. The magenta dotted line shows star counts from the UDS field (which is within the DeepDrill {\em XMM}-LSS field) from \citet{2015ApJS..218...33A}, using the model from \citet{1992ApJS...83..111W} as refined by \citet{1993AJ....105.1860C,1994AJ....107..582C,1995ApJ...444..874C} and \citet{1998ApJ...508...74A}. (As all three fields have Galactic latitudes of $-60^{\circ} \pm 15^{\circ}$ these will be representative of the survey as a whole.) The dashed black and green lines show the galaxy counts from the {\sc Shark} \citep{2018MNRAS.481.3573L,2019MNRAS.489.4196L} and {\sc GALFORM} semi-analytic simulations, respectively, and the dotted red line those from the {\sc EAGLE} hydrodynamic simulations. (b) the same for the 4.5~$\mu$m counts. (c) the survey completeness as a function of magnitude. (d) counts for red ([3.6]$-$[4.5]$>$0.1) and blue ([3.6]$-$[4.5]$<$-0.3) sources separately, with the total [4.5] counts also shown, along with model counts from {\sc Shark}, {\sc GALFORM} and {\sc EAGLE}.}
\label{fig:counts}
\end{figure*}

Source counts from {\em Spitzer} surveys have been presented in \citet{2004ApJS..154...39F}, \citet{2012PASP..124..714M}, \citet{2013ApJS..209...22A}, \citet{2015ApJS..218...33A} and \citet{2018ApJS..237...39A}. We compare the differential counts (number per square degree per magnitude) in the DeepDrill fields (Tables \ref{tab:ch1_counts}, \ref{tab:ch2_counts}) to the S-CANDELS counts of \citet{2015ApJS..218...33A} (the deepest of 
the currently-published post-cryogenic misson counts) in Figure \ref{fig:counts}. Completeness estimates from \citet{2012PASP..124..714M} were used to correct the
counts at AB magnitudes $< 21.25$; at fainter magnitudes we ran a new set of simulations (6000 simulated point sources in each half-magnitude bin) on the DeepDrill ES1 field to improve our estimates of completeness (Figure \ref{fig:counts} (c)). These simulations involved scaling point sources extracted from the mosaics to a known flux density corresponding to the mid-point of the magnitude bin, adding a grid of 6000 of these sources at known positions to the mosaic, and rerunning SExtractor. The resulting catalogues were then matched to the known positions within 1\farcs2, and the fraction of recovered sources noted as the completeness value for that bin. These extractions also allowed us to examine possible biases in the recovered flux densities near the flux density limit from both Eddington bias \citep{1940MNRAS.100..354E} and biases from source confusion resulting in an oversubtracted background. We find significant biases only at the faintest magnitudes -- at $AB=22.75$ (2.9~$\mu$Jy) the recovered flux densities are higher than those input by 3\% at [3.6] and 5\% at [4.5], however, this rises to 15\% at [3.6] and 24\% at [4.5] at $AB=23.25$ (1.8~$\mu$Jy), consistent with Eddington bias dominating. We thus do not list the counts below $AB=22.75$.

The DeepDrill counts agree well with the deeper counts from the literature, starting to diverge slightly at $AB\approx 22$, probably due to  differences in the photometric
algorithms used when the significance of the DeepDrill detections drops below $\approx 10\,\sigma$, and the source confusion limit is approached. 
We also constructed counts from the deeper data in the $\approx 0.45\,$deg$^2$ of the {\em XMM}-LSS field that uses images from the SpUDS survey, verifying that the counts in that area have the same shape as the overall counts to $AB\approx 22.5$, beyond which they are more complete, as expected.

To measure the counts in DeepDrill, we used the flux densities within the 3\farcs8 diameter aperture (with an aperture correction of 0.736 at [3.6] and 0.716 at [4.5]). This will slightly underestimate the fluxes (and hence the counts) at bright magnitudes ($AB \sim 16 - 18$), where the galaxies may be resolved on scales larger than the aperture (at brighter magnitudes the counts are dominated by stars). A comparison to the counts made with the 8\farcs2 diameter aperture (with aperture corrections of 0.920 and 0.905 at [3.6] and [4.5], respectively) shows that the counts in this range are about $\approx 0.1$ dex higher in both bands. Over other magnitude ranges the differences are less than 0.1 dex (except at $AB \stackrel{>}{_{\sim}} 21$, when the higher noise results in more incompleteness in the larger aperture). It is also the case that the aperture corrections are derived from point sources, so both the fluxes and the completeness corrections are strictly incorrect for marginally-resolved galaxies. However, the good agreement between the counts from the two apertures with very different aperture corrections argues against this being a significant problem. The counts of \citet{2015ApJS..218...33A} used a point-source fitting method and agree with ours to within 0.1 dex over the whole magnitude range. The point-source fitting method works better on blended objects, but also can be affected by flux boosting from very low level cosmic rays, whereas our use of SE{\sc xtractor} will probably lead to slight undercounting at faint magnitudes due to blending near the confusion limit.  

\begin{table*}
\caption{Counts at 3.6 $\mu$m and their uncertainties for the three DeepDrill fields, the mean, and those from the simulations.}
\label{tab:ch1_counts}
\begin{tabular}{|c|l|l|l|l|l|l|l|l|l|l|l|l|}
  \multicolumn{1}{|c|}{AB Mag} &
  \multicolumn{1}{c|}{Completeness} &
  \multicolumn{1}{c|}{ES1} &
  \multicolumn{1}{c|}{Error} &
  \multicolumn{1}{c|}{{\em XMM}-LSS} &
  \multicolumn{1}{c|}{Error} &
  \multicolumn{1}{c|}{ECDFS} &
  \multicolumn{1}{c|}{Error} &
  \multicolumn{1}{c|}{Mean} &
  \multicolumn{1}{c|}{Error} &
  \multicolumn{1}{c|}{{\sc Shark}}&
  \multicolumn{1}{c|}{{\sc GALFORM}}&
  \multicolumn{1}{c|}{{\sc EAGLE}} \\
\hline\hline
  14.25 & 1.0 & 1.586 & 0.032 & 1.4 & 0.04 & 1.331 & 0.044 & 1.453 & 0.02 & 0.444&0.521& -\\
  14.75 & 1.0 & 1.928 & 0.022 & 1.883 & 0.023 & 1.91 & 0.022 & 1.907 & 0.013 & 0.785&0.656& -\\
  15.25 & 1.0 & 2.024 & 0.02 & 2.068 & 0.019 & 2.098 & 0.018 & 2.065 & 0.011 & 1.09&0.889 & -\\
  15.75 & 1.0 & 2.185 & 0.016 & 2.157 & 0.017 & 2.229 & 0.016 & 2.191 & 0.009 & 1.387&1.164 & -\\
  16.25 & 1.0 & 2.304 & 0.014 & 2.314 & 0.014 & 2.371 & 0.013 & 2.331 & 0.008 & 1.666& 1.403& -\\
  16.75 & 1.0 & 2.428 & 0.012 & 2.447 & 0.012 & 2.508 & 0.011 & 2.462 & 0.007 & 1.986&1.675&-\\
  17.25 & 1.0 & 2.613 & 0.01 & 2.611 & 0.01 & 2.675 & 0.009 & 2.634 & 0.005 & 2.282&2.023&1.642\\
  17.75 & 1.0 & 2.811 & 0.008 & 2.825 & 0.008 & 2.857 & 0.008 & 2.831 & 0.004 & 2.576&2.337&2.318\\
  18.25 & 1.0 & 3.088 & 0.006 & 3.089 & 0.006 & 3.111 & 0.006 & 3.096 & 0.003 & 2.878&2.658&2.853\\
  18.75 & 1.0 & 3.427 & 0.004 & 3.42 & 0.004 & 3.425 & 0.004 & 3.424 & 0.002 & 3.170&3.018&3.320\\
  19.25 & 0.98 & 3.736 & 0.003 & 3.734 & 0.003 & 3.731 & 0.003 & 3.733 & 0.002 &3.448&3.332&3.668\\
  19.75 & 0.96 & 3.973 & 0.002 & 3.966 & 0.002 & 3.969 & 0.002 & 3.97 & 0.001 & 3.699&3.613&3.915\\
  20.25 & 0.95 & 4.132 & 0.002 & 4.13 & 0.002 & 4.132 & 0.002 & 4.131 & 0.001 & 3.917&3.846&4.111\\
  20.75 & 0.94 & 4.24 & 0.002 & 4.241 & 0.002 & 4.242 & 0.002 & 4.241 & 0.001 & 4.086&4.058&4.238\\
21.25&0.91&4.342&0.008&4.344&0.008&4.345&0.008&4.344&0.008&4.216&4.237&4.338\\
21.75&0.87&4.433&0.008&4.429&0.008&4.435&0.008&4.432&0.008&4.323&4.404&4.417\\
22.25&0.83&4.522&0.009&4.529&0.009&4.523&0.009&4.525&0.009&4.422&4.562&4.494\\
22.75&0.76&4.652&0.010&4.650&0.010&4.646&0.010&4.649&0.009&4.524&4.701&4.580\\
\hline\end{tabular}

\noindent
\begin{flushleft}
Counts and errors are expressed as log$_{10}(N)$, where $N$ is the count per square degree per magnitude. The errors are based on Poisson statistics, combined with the uncertainty in the completeness estimates (which dominate at fainter magnitudes).
\end{flushleft}
\end{table*}

\begin{table*}
\caption{Logarithmic counts at 4.5 $\mu$m and their uncertainties for the three DeepDrill fields, the mean, and those from the simulations.}
\label{tab:ch2_counts}
\begin{tabular}{|c|l|l|l|l|l|l|l|l|l|l|l|l|}
  \multicolumn{1}{|c|}{AB Mag} &
  \multicolumn{1}{c|}{Completeness} &
  \multicolumn{1}{c|}{ES1} &
  \multicolumn{1}{c|}{Error} &
  \multicolumn{1}{c|}{{\em XMM}-LSS} &
  \multicolumn{1}{c|}{Error} &
  \multicolumn{1}{c|}{ECDFS} &
  \multicolumn{1}{c|}{Error} &
  \multicolumn{1}{c|}{Mean} &
  \multicolumn{1}{c|}{Error} &
  \multicolumn{1}{c|}{{\sc Shark}} &
    \multicolumn{1}{c|}{{\sc GALFORM}}&
  \multicolumn{1}{c|}{{\sc EAGLE}} \\
\\
\hline\hline
  14.25 & 1.0 & 1.419 & 0.04 & 1.289 & 0.045 & 1.332 & 0.043 & 1.35 & 0.022 & 0.297&0.085&-\\
  14.75 & 1.0 & 1.761 & 0.027 & 1.825 & 0.024 & 1.859 & 0.023 & 1.817 & 0.014 & 0.604&0.521&-\\
  15.25 & 1.0 & 1.886 & 0.023 & 1.876 & 0.023 & 1.908 & 0.022 & 1.89 & 0.013 & 0.930&0.706&-\\
  15.75 & 1.0 & 2.016 & 0.02 & 2.037 & 0.019 & 2.079 & 0.018 & 2.045 & 0.011 & 1.24&0.919&-\\
  16.25 & 1.0 & 2.201 & 0.016 & 2.176 & 0.016 & 2.252 & 0.015 & 2.211 & 0.009 & 1.545&1.225&-\\
  16.75 & 1.0 & 2.338 & 0.014 & 2.341 & 0.013 & 2.387 & 0.013 & 2.356 & 0.007 & 1.831&1.457&-\\
  17.25 & 1.0 & 2.5 & 0.011 & 2.508 & 0.011 & 2.577 & 0.01 & 2.53 & 0.006 & 2.149&1.803&1.370\\
  17.75 & 1.0 & 2.69 & 0.009 & 2.722 & 0.009 & 2.762 & 0.008 & 2.726 & 0.005 & 2.423&2.149&2.015\\
  18.25 & 1.0 & 2.965 & 0.007 & 2.972 & 0.006 & 2.995 & 0.006 & 2.978 & 0.004 & 2.717&2.453&2.642\\
  18.75 & 1.0 & 3.258 & 0.005 & 3.263 & 0.005 & 3.269 & 0.005 & 3.263 & 0.003 & 3.012&2.776&3.097\\
  19.25 & 0.98 & 3.607 & 0.003 & 3.604 & 0.003 & 3.608 & 0.003 & 3.606 & 0.002 & 3.298&3.116&3.517\\
  19.75 & 0.96 & 3.919 & 0.002 & 3.927 & 0.002 & 3.926 & 0.002 & 3.924 & 0.001 & 3.577&3.450&3.811\\
  20.25 & 0.95 & 4.131 & 0.002 & 4.131 & 0.002 & 4.137 & 0.002 & 4.133 & 0.001 & 3.829&3.755&4.025\\
  20.75 & 0.93 & 4.263 & 0.002 & 4.269 & 0.002 & 4.268 & 0.002 & 4.267 & 0.001 & 4.054&4.015&4.195\\
  21.25 & 0.92 & 4.347 & 0.008 & 4.347 & 0.008 & 4.356 & 0.008 & 4.35 & 0.008 & 4.223&4.224&4.311\\
  21.75 & 0.89 & 4.424 & 0.008 & 4.419 & 0.008 & 4.427 & 0.008 & 4.423 & 0.008 & 4.331&4.394&4.394\\
  22.25 & 0.85 & 4.513 & 0.009 & 4.524 & 0.009 & 4.517 & 0.009 & 4.518 & 0.009 & 4.415&4.546&4.466\\
  22.75 & 0.76 & 4.639 & 0.01 & 4.604 & 0.01 & 4.631 & 0.01 & 4.625 & 0.009 & 4.509&4.665&4.551\\
\hline\end{tabular}

\noindent
\begin{flushleft}
Counts and errors are expressed as log$_{10}(N)$, where $N$ is the count per square degree per magnitude. The errors are based on Poisson statistics, combined with the uncertainty in the completeness estimates (which dominate at fainter magnitudes).
\end{flushleft}

\end{table*}

We also compare our source counts to two semi-analytic simulations, {\sc GALFORM} and {\sc Shark}, and to the Evolution and Assembly of GaLaxies and their Environments ({\sc EAGLE}) hydrodynamic simulations. Semi-analytic simulations can generate large samples for easy comparison to surveys like DeepDrill, but require some assumptions to be made regarding the effects of dust attenuation in relation to the geometry of disks and bulges to compute observed fluxes, compared to hydrodynamical simulations,
which use 3D radiative transfer to calculate these effects directly.

The {\sc GALFORM} simulated counts are based on a lightcone specifically built for SERVS. This SERVS lightcone of model galaxies was constructed using the \citet{2012MNRAS.426.2142L} {\sc GALFORM} model variant using the techniques described in \citet{2013MNRAS.429..556M}. 
\footnote{The data are available from \url{https://zenodo.org/record/3568147\#.X2utDC9h2L6}} 
The lightcone covers the redshift range $0<z<6$, has a sky coverage of $18.09$~deg$^2$ and contains model galaxies with apparent, dust attenuated magnitudes in the [3.6] band down to 2~$\mu$Jy \citep[see also][]{2020ApJ...889..185K}. The {\sc Shark} 
simulated lightcone was built from the {\sc Shark} semi-analytic model of galaxy formation and evolution \citep{2018MNRAS.481.3573L,2019MNRAS.489.4196L} to compare with the DeepDrill survey. This lightcone has an area of $\approx 107.9$~deg$^2$, and a flux selection at the [3.6] band $>0.575\, \mu$Jy (equivalent to an AB magnitude of $24.5$), with the same redshift range of $0\le z\le 6$. 

We computed the predicted number counts from the {\sc EAGLE} hydrodynamical simulations \citep{2015MNRAS.446..521S,2015MNRAS.450.1937C} by extracting all galaxies at $0<z<8$ from their public database \citep{2016A&C....15...72M}, and particularly their redshift, and apparent [3.6] and [4.5] flux densities (see \citet{2018ApJS..234...20C} for details of how these were computed). To calculate the number counts at a given band, we first calculated the area spanned by the box size of $100\,\rm Mpc$ projected in the sky (we assume that the width of the box is negligible with respect to the luminosity distance and hence we can assume all galaxies in the box to be at the same redshift). We then calculate the number of galaxies per unit area at each redshift and apparent magnitude bin. We then integrate the latter over redshift to obtain a total number per unit area in each magnitude bin, which we then divide by the width of the magnitude bin to compare to our other measurements. We do this at $3.6\mu$m, $4.5\mu$m and for subsamples of red and blue IRAC galaxies.

Overall, there is a fair agreement between the models and theobservations in the regime where galaxies dominate the observed counts ($AB\stackrel{>}{_{\sim}}18$). At intermediate magnitudes, $\approx 18$-$21$, the semi-analytic models underpredict the number of galaxies compared to both our counts and those in the literature (by up to a factor of 1.8 in the [3.6] band and 2.2 in the [4.5] band for {\sc Shark}), but the {\sc EAGLE} hydrodynamic simulation does better. At the faint end ($AB\approx 22$) the {\sc GALFORM} model does best, with both {\sc Shark} and {\sc EAGLE} underpredicting the counts. All lightcones contain only the stellar emission of galaxies in the IRAC bands, neglecting warm dust emission from AGNs. Including this may improve the agreement between some of the models and the data.

We also examine the source counts as a function of colour, using the red and blue colour cuts described above to isolate intermediate- ($z\approx 0.7$) and high-redshift ($z\stackrel{>}{_{\sim}}1.3$) galaxies. Figure \ref{fig:counts} (d) shows these counts, again compared
to the model galaxies. Both the blue and red counts start well above all the model counts (in the case of the blue counts at $AB<18$ this is due to stars), with {\sc EAGLE} producing the largest fraction of blue galaxies, and {\sc Shark} the smallest. 
Both sets of observed counts converge to near-agreement with the {\sc Shark} simulation at the faint end, but the other models do less well, with both {\sc EAGLE} and {\sc GALFORM} overpredicting the blue counts, and {\sc GALFORM} also overpredicting the red counts.
This suggests that the redshift distribution of the sources is quite different in the three simulations. With increasing number of photometric and spectroscopic redshift for SERVS and DeepDrill, we will be able to compare the redshift distributions of the data and the simulations, which will add another dimension to the tests shown here. A detailed discussion of the galaxy luminosity function and stellar mass function using DeepDrill data will also be presented in future papers.

\subsection{Very red objects in [3.6]--[4.5]}\label{sec:red}

\begin{figure}
    \centering
    \includegraphics{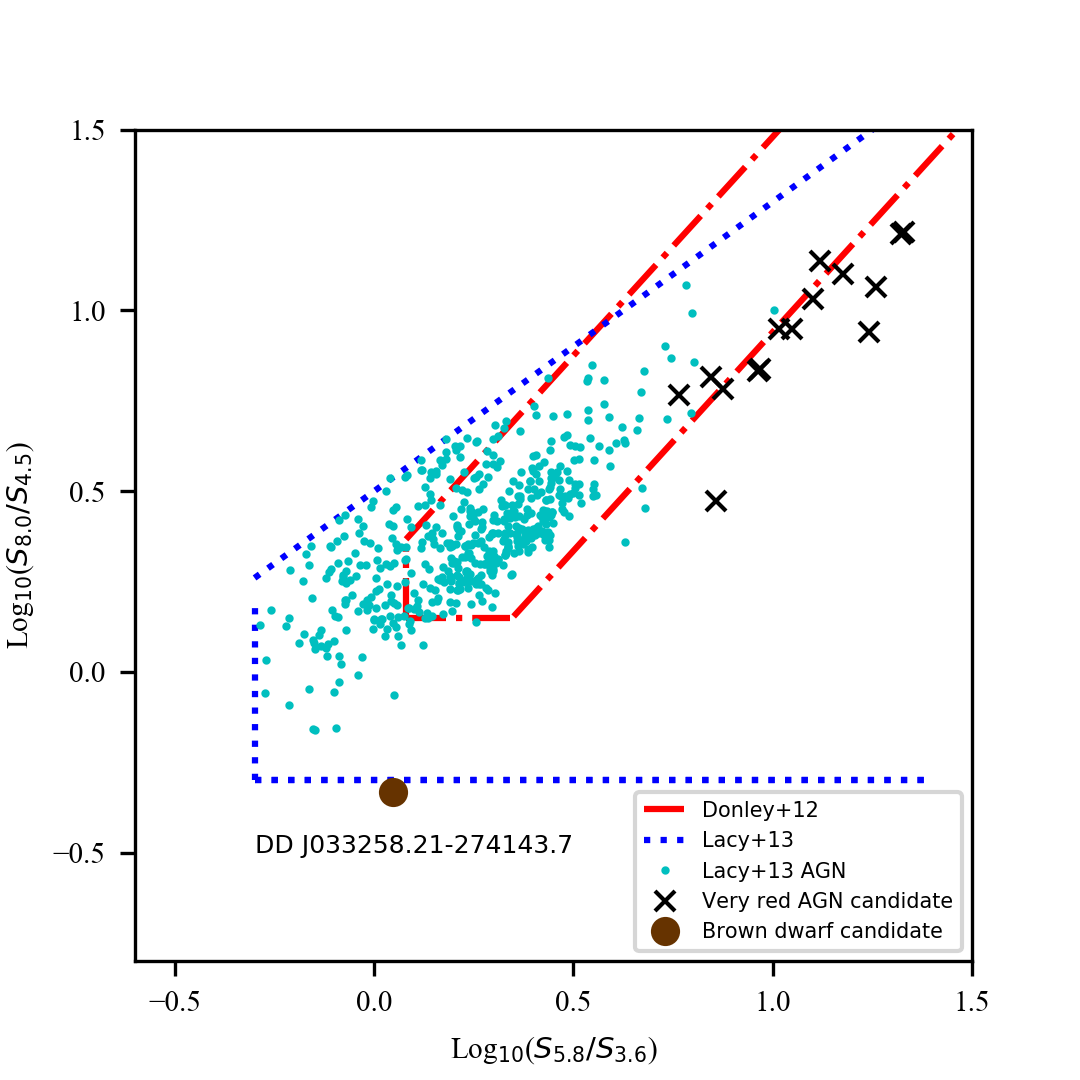}
    \caption{The distribution of the objects with very red [3.6]$-$[4.5] colours in mid-infrared color space for all objects with detections in all four IRAC bands. The black crosses indicate probable AGNs, and the brown dot the likely brown dwarf (DD J033258.21$-$274143.7). The cyan dots represent the spectroscopically-confirmed {\em Spitzer}-selected AGNs from \citet{2013ApJS..208...24L}, with the selection criteria of AGN candidates for that paper shown as the blue dotted line. We also show the AGN selection criterion of \citet{2012ApJ...748..142D} as the red dot-dash line. Both infrared AGN selection techniques shown here rely on the warm dust emission from the AGN outshining the stellar emission from the host galaxy, see \citet{2020NatAs...4..352L} for a detailed discussion.}
    \label{fig:colcol}
\end{figure}

In this section, we investigate the population of objects in DeepDrill at extreme red [3.6]$-$[4.5] colours. Such objects are of interest for both Galactic astronomy, where late-type dwarf stars with strong molecular bands can have extreme [3.6]$-$[4.5] colours \citep{2010ApJ...710.1627L}, and for extragalactic astronomy, where such [3.6]$-$[4.5] colours are sometimes found for highly obscured, highly luminous AGNs \citep{2015ApJ...805...90T}. In Table \ref{tab:very_red} we list the 19 objects detected in both IRAC bands with a signal-to-noise ratio $>3$ ([3.6]$-$[4.5]$>$1.2) that also have 24 $\mu$m coverage in SWIRE (3$\sigma$ depth $\approx 100\mu$Jy; total area of overlap $\approx$ 30 deg$^2$). 16 of the objects have detections at 24 $\mu$m, implying their SEDs continue to rise rapidly in the mid-infrared. These are good candidates for dusty high-$z$ AGN.  
In Figure \ref{fig:colcol} we plot the logarithm of the flux density ratio between 3.6 and 5.8`$\mu$m, $S_{5.8}/S_{3.6}$, versus the logarithm of the flux density ratio between 4.5 and 8.0`$\mu$m, $S_{8.0}/S_{4.5}$ for the 15 of these objects that are covered by the SWIRE IRAC 5.8 $\mu$m and 8.0 $\mu$m observations, and compare to the colours of spectroscopically-confirmed AGNs from  \citet{2013ApJS..208...24L}. Our very red IRAC objects are redder than nearly all of the confirmed AGNs, but follow the extrapolation of the colour trend of AGN to the red \citep[though fall a little below the selection area of][]{2012ApJ...748..142D}. A few of the very red objects have faint detections in VIDEO (Table \ref{tab:very_red}), for these objects the $K_s-$[4.5] colours of the sources detected at 24$\mu$m
are also consistent with AGNs \citep{2012ApJ...754..120M}.

The very red AGNs that are detected at 24 $\mu$m have some similarities to objects selected on the basis of extremely red $R-24\,\mu$m colours in surveys with {\em Spitzer} \citep[dust-obscured galaxies or ``DOGs";][]{2008ApJ...677..943D}, very red
$r-22\,\mu$m colours between the Sloan Digital Sky Survey and {\em WISE} \citep{2015MNRAS.453.3932R}
and objects with very red colours between the {\em WISE} bands
themselves \citep[``Hot DOGs";][]{2015ApJ...805...90T}. These sets of objects are also dominated by AGN emission in the mid-infrared.  However, we emphasize that the [3.6]$-$[4.5] colours of the very red objects presented here are typically more extreme.  None of the DOGs, only 11/77 of the red AGN in \citet{2015MNRAS.453.3932R} and 5/20 of the Hot DOGs have [3.6]$-$[4.5] colours as red as the very red objects described here.

Three sources are not detected at 24 $\mu$m, and are thus less likely to be AGN. One of these (DD J033258.19$-$274143.8) is in the much deeper coverage of S-CANDELS, and has faint detections at 5.8 and 8.0 $\mu$m (Figure \ref{fig:colcol}). This object was also detected in multiple epochs of archival {\em Hubble Space Telescope (HST)} coverage (2004-08-13 using the ACS instrument through the F850LP filter for proposal ID 9500 and 2011-12-30 through F814W for the CANDELS observations, proposal ID 12060). We downloaded the relevant data from the HST archive, and noted that the object has a significant proper motion, moving approximately 0.7 arcsec in R.A. and 0.5 arcsec in Decl. between the two epochs. We therefore tentatively identify this object as a brown dwarf star. With a [3.5]$-$[4.6] colour of 1.6 in AB magnitudes (2.2 in Vega magnitudes), this places the star in a spectral class approximately at the transition between T and Y-dwarfs \citep{2017ApJ...842..118L}. We speculate that the remaining two objects that lack a 24~$\mu$m detection may also be brown dwarfs.

To further investigate the possibility of an AGN origin for the majority of the very red sources we searched for serendipitously available mid-infrared spectroscopy as well as deep radio continuum data in the DeepDrill fields.  
For one of the very red sources, DD J022050.38$-$053714.1 (5MUSES-033), the {\it Spitzer} IRS spectrum from the CASSIS database \citep{2011ApJS..196....8L} shows a strong mid-infrared continuum and spectral features consistent with an AGN at its photometric redshift (derived from SWIRE data) of 2.02 \citep{2013MNRAS.428.1958R}. 
This source is also detected in the VLA Sky Survey (VLASS; \citealt{2020PASP..132c5001L}) with a 3 GHz flux density $S_{\mathrm{3 GHz}}=2.0\pm0.2 \,$mJy. The correspondingly high radio luminosity of $L_{\mathrm{3 GHz}}=5\times$10$^{25}\, {\rm W\; Hz^{-1}}$ is much more consistent with the radio emission being powered by an AGN rather than by star formation \citep[e.g.][]{2011ApJ...739L..29K}. One further source, DD J022008.87-041819.0 is also detected in VLASS with $S_{\mathrm{3 GHz}}=0.54 \pm0.2\,$mJy. Its photometric redshift is 2.01, which implies $L_{\mathrm{3 GHz}}=1.3\times$10$^{25}\, {\rm W\; Hz^{-1}}$, also more likely to be powered by an AGN than by star formation. In addition to VLASS, we also searched for counterparts in the deep radio surveys with published source catalogs listed in Table 4 (LOFAR, GMRT, and ATLAS).  None of the very red sources is detected in these radio surveys.  We also searched for counterparts in deep (sub-mJy~beam$^{-1}$ rms sensitivity) commensal 340~MHz data from the VLA Low-band Ionosphere and Transient Experiment (VLITE\footnote{VLITE provides data at 340~MHz from a subset of the VLA antennas during regular VLA observations above 1~GHz over a field-of-view (measured at the full-width at half power of the primary beam) of 5.5~deg$^2$ and a maximum resolution of $\sim5^{\prime \prime}$ in the VLA A configuration.}; \citealt{clarke+16,polisensky+16}) that were observed during an ultra-deep, single-pointing VLA imaging campaign centered on the {\it Hubble} Ultra Deep Field (HUDF) within CDFS at 3 GHz 
(Abrahams et al. 2020, submitted).  Of the 12 very red sources in CDFS that fall within the currently imaged portion (spanning a width of 2~deg) of the deep VLITE data, one source (DD J033401.66$-$265017.0) has a compact counterpart in VLITE with a primary-beam-corrected flux of 1.0$\pm$0.3~mJy at a SNR of 4.8. The lack of detection in VLASS implies a moderately-steep spectral index, $\alpha^{0.34\,\mathrm{GHz}}_{3.0\,\mathrm{GHz}}<-0.9$. Unfortunately, the source currently lacks a photometric redshift, though assuming it is at $z>1$ it is also more likely to be powered by an AGN. Full details on the deep VLITE imaging centered on the HUDF will be presented in a forthcoming study (Polisensky et al., in preparation).

\begin{table*}[]
    \centering
        \caption{Mid-infrared and radio photometry of the very red sources from DeepDrill with 24 $\mu$m coverage in SWIRE. Infrared photometry is from either the {\em Spitzer} source list \citep[SSTSL2;][]{2010ASPC..434..437T}, this paper, or S-CANDELS, as noted in the Notes column.}
    {\scriptsize
    \begin{tabular}{c|lllllllll}
    Name & $K_s$&$S_{3.6}$ & $S_{4.5}$ & $S_{5.8}$ & $S_{8.0}$ & $S_{24}$ & $S_{\rm 3GHz}$ &Photometric & Notes\\
         & (AB)&($\mu$Jy)& ($\mu$Jy)&($\mu$Jy)&($\mu$Jy)&($\mu$Jy)& (mJy)& Redshift& \\\hline\hline
        DD J003136.66$-$432234.9 &- &1.2$\pm 0.7$ & 6.3$\pm 0.7$ & $<$20& $<20$ &$<$100 &- &-&\\
        DD J003441.85$-$423119.9 &-& 3.8$\pm 0.5$ & 15.2$\pm 0.4$& 81$\pm 4$ & 250$\pm 6$ & 1710$\pm 80$ &- & -&SSTSL2\\
        DD J003449.14$-$422934.8 &-& 2.0$\pm 0.6$ & 6.4$\pm 0.5$& 42$\pm 7$ & 104$\pm 7$ &230$\pm 30$ & -&- & This paper\\
        DD J003723.88$-$432554.7 &-& 1.8$\pm 0.6$ & 6.9$\pm 0.5$& 27$\pm 7$&87$\pm 7$ &290$\pm 30$ &- &- &This paper\\
        DD J003831.45$-$440045.2 &22.39$\pm 0.06$ &3.0$\pm 0.7$ & 10.7$\pm 0.6$& 21$\pm 3$ & 70$\pm 6$ & 390$\pm 80$ &-& - & SSTSL2\\
        DD J004411.11$-$442448.7 &- &2.9$\pm 0.4$ & 9.3$\pm 0.4$ & $<$20 & - &577$\pm 30$&- & & This paper\\
        DD J021821.44$-$053101.5 &-&6.9$\pm 0.5$ &21.4$\pm 1.0$&125$\pm 7$ &248$\pm 7$ &1430$\pm 30$& $<0.4$  &2.02 & \\
        DD J022008.87$-$041819.0 &21.04$\pm 0.02$& 107.6$\pm 3.7$& 337.9$\pm 5.3$ & 997$\pm 30$& 2330$\pm 70$ & 6010$\pm 30$& 0.54$\pm 0.12$ &2.01 &\\
        DD J022017.88$-$045753.3 &-& 4.8$\pm 0.7$ & 15.4$\pm 1.2$  &28$\pm 4$ &90$\pm 8$ & 290$\pm 60$ & $<0.4$ &-& SSTSL2\\
        DD J022050.38$-$053714.1 &- &37.0$\pm 1.3$ & 122.9$\pm 2.5$ & 380$\pm 10$ & 1090$\pm 30$ & 5000$\pm 200$ &2.0$\pm 0.12$& 2.05 & 5MUSES-033; This paper\\
        DD J022228.02$-$034235.3 &-& 43.1$\pm 1.8$ & 143.6$\pm 4.2$ & 320$\pm 10$ & 870$\pm 30$ & 1094$\pm 240$ &$<0.4$ &2.34 & This paper\\
        DD J032727.28$-$270621.9&-&1.8$\pm 0.9$ & 9.7$\pm 0.6$ & $<$20 & $<20$ & $<$100& $<0.15$ & - & \\
        DD J032857.12$-$284111.8&22.80$\pm 0.06$&16.0$\pm 1.2$ & 56.6$\pm 2.2$ & 173$\pm 3$& 443$\pm 5$ &1730$\pm 90$ & $<0.15$& -& SSTSL2, This paper\\
        DD J033029.88$-$293445.5&-&4.5$\pm 0.5$ & 14.5$\pm 0.7$ & 59$\pm 7$ & 198$\pm 7$ &680$\pm 30$ & $<0.15$&- & SSTSL2, This paper\\
        DD J033053.41$-$274717.8&22.41$\pm 0.06$&11.9$\pm 0.8$& 46.7$\pm 2.0$ &150$\pm 5$ & 480$\pm 10$  & 2010$\pm 40$& $<0.15$ &- & SSTSL2\\
        DD J033258.19$-$274143.8&22.93$\pm 0.06$&1.7$\pm 0.5$ & 7.3$\pm 0.9$ & 5.3$\pm 2.2$ & 6.1$\pm 1.8$ &$<$100& $<0.15$& -& S-CANDELS\\
        DD J033400.05$-$283001.9&-&2.7$\pm 0.5$ & 9.2$\pm 0.8$ &47$\pm 3$ & 80$\pm 8$ & 200$\pm 40$ & $<0.15$ &- & SSTSL2\\
        DD J033401.66$-$265017.0&-&24.4$\pm 1.0$ &76.3$\pm 1.7$ & 176$\pm 2$ & 227$\pm 7$& 1000$\pm 50$ &$<0.15$ & -& $S_\mathrm{340 MHz}=$1.2 mJy; SSTSL2, This paper\\
        DD J033602.31$-$284944.2&-&3.6$\pm 0.4$ & 12.8$\pm 1.0$ & 33$\pm 2$ & 87$\pm 5$ & 490$\pm 50$& $<0.15$ & -& SSTSL2\\\hline
    \end{tabular}
    
    \noindent
    \begin{flushleft}
    $^{\dag}$ Photometric redshifts are from \citet{2013MNRAS.428.1958R} except for DD J021821.44$-$053101.5 which is from \citet{2009ApJ...691.1879W}.
    \end{flushleft}
    }
    \label{tab:very_red}
\end{table*}

\section{Summary}\label{sec:summary}

We describe the DeepDrill survey, which images $\approx 27\,$deg$^{2}$ in three of the four pre-defined LSST Deep Drilling fields to a depth of $\approx 2\,\mu$Jy. Accuracy in photometric and astrometric calibration is described. 
We illustrate the use of the [3.6]$-$[4.5] colour to divide objects into high ($z>1.3$) and low ($z\sim 0.7$) redshift bins. This property of the [3.6]$-$[4.5] colour will be particularly valuable for breaking degeneracies in photometric redshift estimates obtained from LSST optical data alone in regions of the DDFs not covered by other deep near-infrared data  \citep[e.g.][]{2013MNRAS.435.1389P}.

We show that source count data at [3.6] and [4.5] alone can provide useful comparisons to models of galaxy formation. The model and observed counts generally agree well, but a small discrepancy observed at intermediate magnitudes warrants further investigation. These
comparisons will be further enhanced when the 
{\em Spitzer} data are fully combined with the other multiwavelength data in the LSST Deep Drilling fields. We are currently working on completing multi-band 0.4-4.5$\mu$m catalogs using forced photometry in the centers of the fields (defined by the overlap between the DeepDrill and VIDEO surveys). We show that most objects with extremely red [3.6]$-$[4.5] colours are mostly identifiable as dusty AGNs at $z>1$, but we also find one likely T or Y brown dwarf star and two further brown dwarf candidates. Future papers will discuss the stellar masses of galaxies in DeepDrill  (Maraston et al.\ 2020, in preparation) and their clustering (van Kampen et al.\ 2020 in preparation).

The 3-5~$\mu$m region in the near-infrared is uniquely useful for capturing light from the peak in the stellar SED at rest frame $\approx 1\mu$m at $z \sim 1-4$. Although new surveys with {\em Spitzer} are no longer possible, we expect that the legacy value of the existing survey datasets will last well into the future. When combined with the existing and planned multiwavlength datasets in these fields, the DeepDrill data will be able to help answer fundamental questions about the nature of AGN and galaxy evolution, and its dependence on environment from before Cosmic Noon to the present.


\section*{Data Availability}

The data products from the post-cryogenic {\em Spitzer} surveys of the three LSST DDFs described here (images, coverage maps, uncertainty images, bright star masks and single and dual-band catalogues) are available from IRSA (\url{https://irsa.ipac.caltech.edu/data/SPITZER/DeepDrill}). These include mosaic images, coverage maps, uncertainty images and bright star masks. Each field has two single-band catalogues cut at $5\sigma$, and a dual-band catalogue requiring a detection at $>3\sigma$ at both 3.6 and 4.5$\mu$m. The simulated lightcone catalogue from {\sc SHARK} is also included in the release, its columns are described in Table \ref{tab:shark}.

\begin{table}
    \caption{Columns in the {\sc SHARK} simulated lightcone catalogue.}
    \centering
    \begin{tabular}{l|ll}
       Column(s)  & Description & Units \\\hline\hline
        1 & ID & -\\
        2 & Right Ascension & Degrees\\
        3 & Declination & Degrees\\
        4 & Redshift &-\\
        5 & Log$_{10}$(Stellar Mass) & Solar masses\\
        6 & Log$_{10}$(Star Formation Rate) & Solar masses/year\\
        7 & Half-light radius & Kpc\\
        8 & Bulge-to-Total ratio & -\\
        9-13 & Apparent\ mag.\ in $u,g,r,i,z$ & AB\\
        14-17 & Apparent mag.\ in $Y, J, H, K$ (VISTA) & AB\\
        18 & Apparent mag.\ in WISE [3.4] & AB\\
        19-20&Apparent mag.\ in IRAC [3.6] and [4.5] & AB\\
        21 & Apparent mag.\ in WISE [4.6]& AB\\
        22-34& Absolute mag.\ in the above bands & AB\\\hline
    \end{tabular}
    \label{tab:shark}
\end{table}

\section*{Acknowledgements}
We would like to thank the referee for a thorough review that significantly improved the paper. We also thank the contributors to the original DeepDrill proposal not listed as coauthors on this paper, including M.\ Dickinson, I.\ Prandoni and L.\ Spitler, and especially I.\ Smail for helpful comments and suggestions. This work is based on observations made with the Spitzer Space Telescope, which was operated by the Jet Propulsion Laboratory, California Institute of Technology under a contract with NASA. Support for this work was provided by NASA through an award issued by JPL/Caltech.
The National Radio Astronomy Observatory is a facility of the National Science 
Foundation operated under cooperative agreement by Associated Universities, Inc. WNB acknowledges support from NASA grant 80NSSC19K0961.
GW acknowledges support from the National Science Foundation through grant AST-1517863, by HST program number GO-15294, and by grant number 80NSSC17K0019 issued through the NASA Astrophysics Data Analysis Program (ADAP).
Support for program number GO-15294 was provided by NASA through a grant from the Space Telescope Science Institute, which is operated by the Association of Universities for Research in Astronomy, Incorporated, under NASA contract NAS5-26555. 
IRS acknowledges support from the United Kingdom Science and Technology Funding Council (ST/P000541/1). Basic research in radio astronomy at the Naval Research Laboratory is funded through 6.1 Base funding. This research has made use of the NASA/IPAC Extragalactic Database (NED), which is operated by the Jet Propulsion Laboratory, California Institute of Technology,
under contract with the National Aeronautics and Space Administration.
This work made extensive use of {\sc topcat} \citep{2011ascl.soft01010T} for catalog matching
and analysis. 

\bibliographystyle{mnras}
\bibliography{deepdrill}







\appendix

\section{Source confusion in the DeepDrill survey}\label{sec:confusion}

To better understand how source confusion affects the DeepDrill survey as a function of depth, and in the presence of source clustering, we make use of some results from studies at other wavelengths. The analytic approach described here, although only approximate, allows us to predict trends in the noise in the survey as a function of depth, and to compare the contributions of random and clustered sources to the confusion noise. 

Source confusion and its contribution to flux and number count uncertainties has been investigated by several authors \citep[e.g.][]{1957PCPS...53..764S,1974ApJ...188..279C,1987ApJS...63..311H,
2012ApJ...758...23C,2016MNRAS.462.2934V}. 
\citet{1987ApJS...63..311H} show that the confusion noise in the absence of clustering, $\sigma_{\mathrm{random}}$, obeys:
\begin{equation}
\label{eqn:confusion}
\sigma_{\mathrm{random}}^2=\frac{\alpha}{2-\alpha} \Omega_{\mathrm e} A D_{\mathrm c}^{2-\alpha},
\end{equation}
where $\alpha$ is the slope and $A$ the normalization of the integrated source counts $n(>S)=A S^{-\alpha}$ near the survey limit, $D_{\mathrm c}$ is the cutoff in the ``deflection'' - the deviation of flux density, $S$ from the mean background - at the survey limit, and the effective source area $\Omega_{\mathrm e}$ is given by:
\begin{equation}
\label{eqn:omega}
\Omega_{\mathrm e}=\int h(\boldsymbol{x})^{\alpha} {\mathrm d}\boldsymbol{x},
\end{equation}
where $h(\boldsymbol{x})$ represents the response of the instrument to a typical source near the survey limit as a function of position vector $\boldsymbol{x}$.
For IRAC, we use the extended Point Response Functions (PRFs) from the {\em Spitzer} Science Center\footnote{\url{https://irsa.ipac.caltech.edu/data/SPITZER/docs/irac/calibrationfiles/psfprf/}} as an empirical $h(\boldsymbol{x})$. The slope of the integrated source counts close to the flux density limit in both [3.6] and [4.5] is $\alpha\approx 0.5$. The non-Gaussian wings of the IRAC PRF, which are further amplified relative to the core by taking the PRF to the 0.5 power in Equation \ref{eqn:omega}, means that the integral is slow to converge, and we chose to limit the integration to the inner 1/4 of the IRAC array where the bulk of the contributions will occur. This integration results in effective area at [3.6] of $\Omega_{e}=1.82\times 10^{-9}$sr, and at [4.5] of $\Omega_{e}=1.86\times 10^{-9}$sr. These values are much larger than that of the 1\farcs8 FWHM Gaussian that corresponds to the core of the IRAC PRF at these wavelengths ($\Omega_{\mathrm{e}} =1.7\times 10^{-11}$sr), and are also much larger than the solid angle of the 1\farcs9 radius aperture through which the source counts are measured ($\Omega_{\mathrm{e}}=2.7\times 10^{-10}$sr). 

Galaxy clustering also adds to confusion noise. The effects of clustering on the distribution of deflections, $P(D)$, have been discussed by \citet{1992ApJ...396..460B}, \citet{2004ApJ...604...40T} and \citet{2019PASP..131h4101A}, and a full treatment requires complicated mathematics outside the scope of this paper. Here, we adopt the approach of \citet{1992ApJ...396..460B}, where we sum the confusion noise contribution from unclustered sources and a term due to clustering in quadrature to estimate the total confusion noise $\sigma_\mathrm{conf}$:
\begin{equation}
    \sigma^2_{\mathrm{conf}}=\sigma^{2}_{\mathrm{random}}+<I>^{2}\Xi,
\end{equation}
where $<I>$ is the mean intensity per beam area, $\Omega_{\mathrm b}$, and 
\[\Xi = \frac{1}{\Omega_{\mathrm b}^2}\int_{\Omega_{\mathrm b}} d\boldsymbol{x_1}\int_{\Omega_{\mathrm b}} w(|\boldsymbol{x_1}-\boldsymbol{x_2}|) \mathrm{d} \boldsymbol{x_2} \]
where w($\theta$) (and $\theta=|\boldsymbol{x_1}-\boldsymbol{x_2}|$) is the two-point correlation function.
Measurements of the two-point correlation function in DeepDrill (van Kampen et al.\ in preparation) show that $w(\theta)=(\theta/\theta_0)^{-\delta}$ where $\delta\approx 0.68$, and the correlation length, $\theta_0\approx0\farcs2$. 

We next approximate the total noise $\sigma_{\mathrm{T}}$ as the contribution of instrumental noise, $\sigma_{\mathrm{I}}$ (obtainable from the {\em Spitzer} Performance Estimation Tool \footnote{\url{https://irsa.ipac.caltech.edu/data/SPITZER/docs/dataanalysistools/tools/pet/senspet/index.html}}) and confusion noise, $\sigma_{\mathrm{T}}=\sqrt{\sigma_{\mathrm{I}}^2+\sigma_{\mathrm{conf}}^2}$. (With the caveat that the distribution of deflections $P(D)$ is not Gaussian; \citet{2004A&A...424.1081H}.) By setting $D=\sigma_{\mathrm{T}}$ in Equation \ref{eqn:confusion} and solving numerically, we recover $\sigma_\mathrm{T}=0.41$~$\mu$Jy at [3.6] and $0.51$~$\mu$Jy at [4.5] at the nominal survey depth of 12$\times$100s frames, close to the empirical result from SERVS of $\sigma_{\mathrm{T}}\approx 0.4\;\mu$Jy in both the [3.6] and [4.5] bands found from measuring the RMS variation in randomly-placed 1\farcs9 radius empty apertures \citep{2012PASP..124..714M}. Table \ref{tab:depth} shows that, as the depth of the survey is increased beyond the nominal depth of 12 frames, $\sigma_\mathrm{T}$ decreases, but by factors less that the square root of the number of frames, and the contribution of confusion noise becomes dominant. The contribution from clustering is small, but almost constant with depth as it is dominated by bright sources well above the survey limit, so becomes more significant in deeper surveys. Source confusion in very deep IRAC images can be effectively mitigated by PRF subtraction or forced photometry techniques \citep[e.g.][]{2015ApJS..218...33A, 2017ApJS..230....9N} that remove the PRF wings from the images, reducing $\Omega_\mathrm{e}$ and $\Omega_\mathrm{b}$.

Positional accuracy is also affected by source confusion \citep{2001AJ....121.1207H}. In the case of DeepDrill, however, the low value of $\alpha$ results in the effect of centroid errors due to source blending at the measured $\approx 30$ beams per source being $\stackrel{<}{_{\sim}}$0\farcs1 \citep[][their Figure 2]{2001AJ....121.1207H}.

\begin{table*}
    \caption{Calculated noise as a function of depth in DeepDrill}
    \centering
    \begin{tabular}{c|cccccccccc}
    Depth         & \multicolumn{2}{c}{$\sigma_\mathrm{I}$} & \multicolumn{2}{c}{$\sigma_\mathrm{random}$}& \multicolumn{2}{c}{$<I> \sqrt\Xi$} & \multicolumn{2}{c}{$\sigma_\mathrm{conf}$}&\multicolumn{2}{c}{$\sigma_\mathrm{T}$}\\
     (100s    & \multicolumn{2}{c}{($\mu$Jy)} &\multicolumn{2}{c}{($\mu$Jy)}&\multicolumn{2}{c}{$(\mu$Jy)}&\multicolumn{2}{c}{$(\mu$Jy)}& \multicolumn{2}{c}{($\mu$Jy)} \\
     frames)& [3.6]&[4.5]&[3.6]&[4.5]&[3.6]&[4.5]&[3.6]&[4.5]& [3.6]&[4.5]\\\hline
     6 & 0.35 & 0.51& 0.33& 0.41 & 0.15 & 0.13 &0.36&0.43&0.51 & 0.67\\ 
     9 & 0.29 & 0.42 & 0.30 & 0.36 & 0.15 & 0.13 &0.34&0.38& 0.45 & 0.57\\ 
     12 &0.25 &0.36 &0.28 & 0.33 & 0.15 & 0.14 &0.32&0.36&0.41 & 0.51\\ 
     24 &0.18 & 0.25 & 0.25 & 0.28 & 0.16 & 0.14 &0.30&0.31& 0.34 & 0.41\\ 
     48 &0.13 & 0.18 & 0.23 & 0.24 & 0.16 & 0.14 &0.28&0.28& 0.30 & 0.33\\ 
     96 &0.09 & 0.13 & 0.21 & 0.22 & 0.16 & 0.14 &0.27&0.26& 0.28 & 0.29\\
    \end{tabular}
    \label{tab:depth}
\end{table*}

\medskip
\noindent
\section{Author affiliations}
\begin{flushleft}
\textit{
$^{1}$National Radio Astronomy Observatory, 520 Edgemont Road, Charlottesville, VA 22903, USA\\
$^{2}$Infrared Processing and Analysis Center, MS 100-22, California Institute of Technology, Pasadena, CA 91125, USA\\
$^{3}$Department of Physics and Astronomy, University of Hawaii, 2505 Correa Road, Honolulu, HI 96822,USA\\
$^{4}$Institute for Astronomy, 2680 Woodlawn Drive, University of Hawaii, Honolulu, HI 96822, USA\\
$^{5}$National Research Council, resident at the Naval Research Laboratory, Washington, DC 20375, USA\\
$^{6}$Instituto de Astrof\'{i}sica e Ci\^{e}ncias do Espa\c co, Universidade de Lisboa, OAL, Tapada da Ajuda, PT1349-018 Lisboa, Portugal\\
$^{7}$Departamento de F\'{i}sica, Faculdade de Ci\^{e}ncias, Universidade de Lisboa, Edif\'{i}cio C8, Campo Grande, PT1749-016 Lisbon, Portugal\\
$^{8}$Department of Astronomy and Astrophysics, 525 Davey Lab, The Pennsylvania State University, University Park, PA 16802, USA\\
$^{9}$Institute for Gravitation and the Cosmos, The Pennsylvania State University, University Park, PA 16802, USA\\
$^{10}$Department of Physics, 104 Davey Lab, The Pennsylvania State University, University Park, PA 16802, USA\\
$^{11}$Astrophysics Group, Imperial College London, Blackett Laboratory, Prince Consort Road, London SW7 2AZ, UK\\
$^{12}$International Centre for Radio Astronomy Research (ICRAR), University of Western Australia,
35 Stirling Hwy, Crawley, WA 6009, Australia.\\
$^{13}$ARC Centre of Excellence for All Sky Astrophysics in 3 Dimensions (ASTRO 3D)\\
$^{14}$Cosmic Dawn Center (DAWN), Copenhagen, Denmark\\
$^{15}$Institute of Cosmology and Gravitation, Dennis Sciama Building, Burnaby Road, Portsmouth PO1 3FX, UK\\
$^{16}$ESA/ESTEC Keplerlaan1, 2201 AZ, Noordwijk, The Netherlands\\
$^{17}$Department of Physics and Astronomy, Tufts University, 212 College Avenue, Medford, MA 02155, USA\\
$^{18}$Department of Physics and Astronomy, University of Pennsylvania, Philadelphia, PA 19104, USA\\
$^{19}$Department of Physics and Astronomy, University of the Western Cape, Private Bag X17, Bellville 7535, South Africa.\\
$^{20}$INAF - Istituto di Radioastronomia, via Gobetti 101, I-40129 Bologna, Italy\\
$^{21}$Department of Physics and Astronomy, University of California, Riverside, 900 University Avenue, Riverside, CA 92521, USA\\
$^{22}$Center for Relativistic Astrophysics, School of Physics, Georgia Institute of Technology, 837 State Street, Atlanta GA 30332-0430, USA\\
$^{23}$Department of Physics and Astrophysics, University of North Dakota, Grand Forks, ND 58202-7129, USA\\
$^{24}$Naval Research Laboratory Remote Sensing Division, Code 7213, 4555 Overlook Avenue SW, Washington, DC 20375, USA\\
$^{25}$Department of Physics and Astronomy, University of California, 4129 Reines Hall, Irvine, CA 92697, USA\\
$^{26}$Department of Physics, University of Napoli "Federico II", via Cinthia 9, 80126 Napoli, Italy\\
$^{27}$Department of Physics and Michigan Center for Theoretical Physics, University of Michigan, Ann Arbor, MI 48109, USA \\
$^{28}$Department of Astronomy, University of Michigan, Ann Arbor, MI 48109, USA \\
$^{29}$Space Telescope Science Institute, 3700 San Martin Drive, Baltimore, MD 21218, USA\\
$^{30}$ Kavli Institute for Cosmological Physics, University of Chicago, Chicago, IL 60637, USA\\
$^{31}$ Fermi National Accelerator Laboratory, PO Box 500, Batavia, IL 60510, USA\\
$^{32}$Astrophysics Research Institute, Liverpool John Moores University, 146 Brownlow Hill, Liverpool L3 5RF, UK\\
$^{33}$ European Southern Observatory, Karl-Schwarzschild-Str 2, D-85748 Garching bei München, Germany\\ 
$^{34}$ National Astronomical Observatories of China, Chinese Academy of Sciences, Beijing 100012, China\\
$^{35}$ China-Chile Joint Center for Astronomy, Chinese Academy of Sciences, Camino El Observatorio, 1515 Las Condes, Santiago, Chile\\
$^{36}$ Center for Astrophysics $\mid$ Harvard and Smithsonian, 60 Garden Street, Cambridge, MA 02138, USA\\
$^{37}$ National Radio Astronomy Observatory, 1003 Lopezville Road, Socorro, NM 87801, USA\\
$^{38}$Astrophysics, Department of Physics, Keble Road, Oxford OX1 3RH, UK\\
$^{39}$ University of Kansas, Department of Physics and Astronomy, 1082 Malott, 1251 Wescoe Hall Drive, Lawrence, KS 66045\\
$^{40}$Australian Astronomical Observatory, North Ryde, NSW 2113, Australia\\
$^{41}$ Department of Physics, University of California-Davis, One Shields Avenue, Davis, CA 95616, USA\\
$^{42}$ Department of Astronomy, University of Cape Town, Private Bag X3, Rondebosch 7701, Cape Town, South Africa\\
$^{43}$ Center for Cosmology and Astro-Particle Physics, The Ohio State University, Columbus, OH 43210, USA\\
$^{44}$ Department of Astronomy, The Ohio State University, Columbus, OH 43210, USA\\
$^{45}$Institute of Astronomy, Madingley Road, Cambridge CB3 0HA, UK\\
$^{46}$Universit\'{e} de Paris, F-75013, Paris, France, LERMA, Observatoire de Paris, PSL Research University, CNRS, Sorbonne Universit\'e,  F-75014 Paris, France\\
$^{47 }$ Jet Propulsion Laboratory, Cahill Center for Astronomy \& Astrophysics, California Institute of Technology, 4800 Oak Grove Drive, Pasadena, CA, USA\\
$^{48}$ Joint ALMA Observatory, Alonso de Córdova 3107, Vitacura 763-0355, Santiago, Chile\\
$^{49}$ Department of Physics and Astronomy and the Pittsburgh Particle Physics, Astrophysics and Cosmology Center (PITT PACC), University of Pittsburgh, Pittsburgh, PA 15260, USA\\
$^{50}$ Western Sydney University, Locked Bag 1797, 1797, Penrith South, NSW, Australia\\
$^{51}$ Astronomy Centre, Department of Physics and Astronomy, University of Sussex, Brighton BN1 9QH, UK\\
$^{52}$ Instituto de Astrof\'{i}sica de Canarias (IAC), Calle V\'{i}a Lactea s/n, E-38205 La Laguna, Tenerife, Spain\\
$^{53}$ Universidad de La Laguna, Dpto. Astrof\'{i}sica, E-38206 La Laguna, Tenerife, Spain\\
$^{54}$ AIM, CEA, CNRS, Universit\'{e} Paris-Saclay, Universit\'{e} Paris Diderot, Sorbonne Paris Cit\'{e}, 91191, Gif-sur-Yvette, France\\
$^{55}$Department of Physics, Drexel University, 32 S. 32nd Street, Philadelphia, PA 19104 USA\\
$^{56}$NSF's National Optical-Infrared Astronomy Research Laboratory, 950 N.\ Cherry Avenue, Tucson, AZ 85719, USA\\
$^{57}$Leiden Observatory, Leiden University, PO Box 9513, NL-2300 RA Leiden, the Netherlands\\
$^{58}$International Centre for Radio Astronomy Research, Curtin University, Bentley, WA 6102, Australia\\
$^{59}$Center for Computational Astrophysics, Flatiron Institute, 162 Fifth Avenue, New York, NY 10010, USA; Department of Physics and Astronomy, Rutgers, The State University of New Jersey, 136 Frelinghuysen Rd, Piscataway, NJ 08854, USA\\
$^{60}$Department of Astrophysical Sciences, Princeton University, Princeton, New Jersey 08544, USA\\
$^{61}$George P. and Cynthia Woods Mitchell Institute for Fundamental Physics and Astronomy, Department of Physics and Astronomy, Texas A \& M University, 4242 TAMU, College Station, TX 77843, USA\\
$^{62}$Haverford College, Departments of Physics and Astronomy, 370 Lancaster Avenue, Haverford, PA 19041, USA\\
$^{63}$Kavli Institute for Particle Astrophysics and Cosmology and Department of Physics, Stanford University, Stanford, CA 94305, USA; Department of Particle Physics and Astrophysics, SLAC National Accelerator Laboratory, Stanford, CA 94305, USA\\
$^{64}$Pittsburgh Particle Physics, Astrophysics, and Cosmology Center (Pitt-PACC), University of Pittsburgh, Pittsburgh, PA 15260, USA}
\end{flushleft}



\bsp	
\label{lastpage}
\end{document}